\documentclass[12pt]{iopart}
\usepackage{iopams,bm,cite,graphicx}
\expandafter\let\csname equation*\endcsname\relax
\expandafter\let\csname endequation*\endcsname\relax

\usepackage{amsfonts}
\usepackage{amsmath}
\usepackage{amsthm}
\usepackage{amssymb}
\usepackage{relsize}
\usepackage[skip=15pt,font=small]{caption}

\newtheorem{theorem}{Theorem} 
\newtheorem{proposition}[theorem]{Proposition}

\usepackage{bbold,dsfont}
\usepackage{upgreek,mathrsfs}
\usepackage[dvipsnames]{xcolor}

\newcommand{\ket}[1]{|#1\rangle}
\newcommand{\bra}[1]{\langle#1|}
\newcommand{\amp}[2]{\langle#1|#2\rangle}
\newcommand{\vect}[1]{\bm{#1}}
\newcommand{\ov}[1]{\overline{\vect{#1}}}
\renewcommand{\Tr}{\mathrm{Tr}}
\DeclareMathOperator{\Sym}{Sym}
\DeclareMathOperator{\Diag}{Diag}
\newcommand{\Mult}{\mathcal{M}}
\newcommand{\one}{\mathbf{1}}

\newcommand{\margvs}{  {\mathlarger\rho}_{\vect{\nu},\vect{\Gamma}}  }
\newcommand{\margvsn}{  {\mathlarger\rho}_{\ov{n},\vect{\Gamma}}  }
\newcommand{\ii}{\mathrm{i}}

\begin{document}

\title[Semiclassical asymptotics of multiphotonic scattering probabilities]{Semiclassical asymptotics of multiphotonic scattering probabilities with partial indistinguishability}

\author{Alonso Botero$^{1}$,   Hubert~de~Guise$^{2}$ \footnote{to whom correspondence should be addressed}, and Orlando Melo-\!Vargas$^{1}$}

\address{
$^1$Departamento de F\'isica, Universidad de los Andes, Bogot\'a, Colombia. \newline 
$^2$Department of Physics, Lakehead University, Thunder Bay, ON, P7B 5E1, 
Canada.}

\begin{abstract}
We propose a framework for computing multiphotonic scattering probabilities in a lossless multiport interferometer for arbitrary photon numbers and degrees of indistinguishability. By exploiting a toroidal expansion of multiphotonic states in tensor powers of single-particle states, the framework defines a map from a torus of relative phases to the probability simplex that governs the asymptotic behavior of scattering probabilities in the large-photon limit. Specifically, the probabilities concentrate on the ``classically allowed region" defined by the map, and the slowly-varying part of the multiphotonic distribution reproduces a classical measure induced by the map. As a result, we are able to establish a new asymptotic formula for the multiphotonic probabilities in a general scenario of partially indistinguishable photons, while also providing a single-particle picture to explain the asymptotics of known multiphotonic transition amplitudes in the fully indistinguishable case. More broadly,  our framework yields  new,   directly testable consequences in relation to asymptotic photon bunching patterns: it translates features of the classical map — such as caustics and voids — into direct predictions about regions of large or exponentially suppressed photon-distribution probability.

\end{abstract}

\maketitle

\section{Introduction}

Multiphotonic interference is a quantum phenomenon of considerable fundamental 
and practical interest. On the fundamental side, it provides a vivid manifestation 
of genuine quantum interference of multiparticle probability amplitudes 
\cite{hongMeasurementSubpicosecondTime1987,mandelQuantumEffectsOnephoton1999,tichyInterferenceIdenticalParticles2014}, 
with no classical counterpart in the statistics of independent particles; on the 
practical side, it underpins applications from metrology to quantum information 
processing 
\cite{carolanUniversalLinearOptics2015,KokLovettIntroductiontoOptica2014l,hamerlyStabilitySelfConfiguringLarge2022} 
and optical neural networks \cite{shenDeepLearningCoherent2017}. The deployment 
of increasingly large linear multiport interferometers has made the evaluation of 
multiphotonic scattering probabilities a problem in which fundamental and practical 
considerations converge. For perfectly indistinguishable photons, the theory is 
elegant: the $N$-photon space is the symmetric subspace $\Sym^N(\mathbb{C}^M)$ of 
the full $N$-particle Hilbert space, and detection probabilities reduce to squared 
matrix permanents \cite{caianielloQuantumFieldTheory1953,englComplexScatteringCanonical2015}, 
which are themselves but one of a series of functions defined over the group 
$GL(M,\mathbb{C})$ \cite{LittlewoodGroupCharacters1934}, selected by the input 
and output photon distributions \cite{spivakGeneralizedInterferenceFermions2022}. 
The elegance of this description, however, comes at a computational cost: under 
reasonable assumptions, permanents of submatrices of sufficiently large random 
unitary matrices are believed to be hard to evaluate 
\cite{valiantComplexityComputingPermanent1979}; it is precisely this hardness 
that underlies the proposal of boson sampling 
\cite{aaronsonComputationalComplexityLinear2010} as a demonstration of quantum 
computational advantage.

Given this computational difficulty, asymptotic methods that capture the behavior 
of multiphotonic scattering probabilities for large photon numbers, without 
requiring the  evaluation of permanents, are of particular interest. In the 
regime where the number of ports $M$ is fixed and the total photon number $N$ is 
large, a select body of work exists for the case of perfectly indistinguishable 
photons. The earliest results concern the two-port beam splitter ($M=2$), where 
the connection with $SU(2)$ group functions 
\cite{yurkeSU2SU11Interferometers1986,camposQuantummechanicalLosslessBeam1989} 
was used to derive a semiclassical picture for the envelope of the output 
distribution; the asymptotics of $SU(2)$ and $SU(3)$ group functions have also 
been explored in other contexts 
\cite{brussaardClassicalLimitsClebschgordan1957a,roweAsymptoticLimitsSU22001,deguiseAsymptoticSU2SU32004,braunSemiclassicsRotationTorsion1996}. 
More recently, semiclassical formulas for the transition amplitudes of general 
$M$-port interferometers 
\cite{shchesnovichASYMPTOTICEVALUATIONBOSONIC2013,englComplexScatteringCanonical2015} 
have provided the important extension to general multiport multiphoton 
interferometry with perfectly indistinguishable photons.

Real photons, however, are often---by design or as a result of experimental 
limitations---only \textit{partially} indistinguishable, differing in internal 
degrees of freedom such as polarization, frequency, or temporal profile. 
Partial distinguishability interpolates between coherent multiphoton 
interference and the statistics of classical, distinguishable particles. Its 
systematic understanding has become increasingly important for benchmarking 
quantum advantage, designing scalable photon sources and interferometers, and 
defining realistic performance thresholds for quantum technologies 
\cite{spivakGeneralizedInterferenceFermions2022,hovenMultiphotonInterferenceOutperforms2025,correaanguitaExperimentalValidationBoson2025,gellerMeasuringMultiparticleIndistinguishability2026,shchesnovichPartialIndistinguishabilityTheory2015,tichySamplingPartiallyDistinguishable2015,dufourFourierAnalysisManybody2024}.
With the commissioning of large interferometers, the question of asymptotics 
of partially distinguishable photons has come to the fore 
\cite{tichyInterferenceIdenticalParticles2014}. Exact expressions for the 
scattering probabilities are available 
\cite{shchesnovichPartialIndistinguishabilityTheory2015,tichySamplingPartiallyDistinguishable2015}, 
but the correlation between port and internal degrees of freedom makes them 
far less amenable to asymptotic analysis than in the indistinguishable case, 
and results in the large-$N$ regime remain limited to the two-port 
interferometer \cite{villalobosQuantumBeamSplitter2023}. Extending this 
analysis to general multiport interferometers is the motivation of this paper.

Here, our aim is to extend the asymptotic theory of multiphotonic scattering 
probabilities to partially indistinguishable photons in a general $M$-port 
interferometer, through a framework that treats all degrees of 
indistinguishability on the same footing, yields directly testable 
predictions about the multiphotonic distribution, and provides a transparent 
physical interpretation of the asymptotic results. In particular, the 
framework should seamlessly recover known asymptotic results for perfectly 
indistinguishable photons 
\cite{englComplexScatteringCanonical2015,shchesnovichASYMPTOTICEVALUATIONBOSONIC2013} 
as a  limiting case.

We consider a class of multiphotonic states we call \emph{correlated mode 
states}, in which all photons entering a given port share one internal state, 
but photons in different ports may have partially overlapping internal modes; 
such states are fully characterized by the input occupations and the Gram 
matrix $\vect{\Gamma}$ of the internal overlaps. By expanding correlated mode 
states in tensor powers (i.e., $SU(M^2)$-coherent states) of single-particle ``virtual states'' parametrized 
over a torus of relative phases (Propositions 1 and 2), we obtain an 
integral representation of the scattering probabilities from which a 
classical map---from the torus to the probability simplex of output 
intensities---emerges naturally. This map underpins the main results of the 
paper: the scattering probabilities concentrate on the classically allowed 
region defined by the image of the map (Theorem 3); the slowly-varying part 
of the distribution converges asymptotically to a classical measure induced 
by the map (Theorem 4); and the rank of $\vect{\Gamma}$ determines whether 
rapidly oscillating interference fringes survive in the large-$N$ limit 
(Theorem 5).

To our knowledge, this is the first derivation of asymptotic formulas for 
the multiphotonic scattering probabilities of partially indistinguishable 
photons in a general multiport interferometer. A notable consequence is 
that for full-rank $\vect{\Gamma}$, the asymptotic distribution reduces 
 to the classical measure, which for interferometers with a small 
number of ports is straightforward to compute. Moreover, the classical map 
reveals a rich structure in the multiphotonic distribution that has no 
counterpart in the two-port case: regions of exponentially suppressed 
probability, which we call \emph{voids}, arising from the topology of the 
map, and ridges of enhanced probability at the singularities of the map, 
which we call \emph{caustics}. These features are already clearly 
identifiable at moderate photon numbers, as illustrated in 
Figs.~\ref{fig:dist} and \ref{fig:indist} for a three-port interferometer.

The paper is organized as follows. Section~\ref{sec:setup} fixes notation, 
defines the multiphotonic scattering probability for a general $M$-port 
interferometer, and introduces correlated mode states as a natural family of 
partially indistinguishable input states. Section~\ref{sec:vs} develops the 
main technical tool: the tensor-power representation of correlated mode 
states in terms of single-particle states---which we term \emph{virtual 
states}---the toroidal expansion over a torus of relative phases 
(Propositions 1 and 2), and the resulting integral representation of the 
scattering probability, from which the classical map and the relevant 
contributions to the large-$N$ limit are identified. Section~\ref{sec:asy} 
carries out the asymptotic analysis: it introduces the classical map and the 
classical measure, establishes the concentration of the scattering 
probability on the image of the classical map (Theorem 3), derives the 
asymptotic form of the slowly-varying density (Theorem 4), proves the 
suppression of interference terms for full-rank $\vect{\Gamma}$ (Theorem 5), 
and discusses the semiclassical formula for the transition amplitudes in the 
limit of perfectly indistinguishable photons. Finally, section~\ref{sec:bunching} 
discusses the observable consequences of the asymptotic results---voids and 
caustics in the multiphotonic distribution---illustrated for a three-port 
interferometer.

\begin{figure}[ht]
\centering
 \includegraphics[scale=0.45]{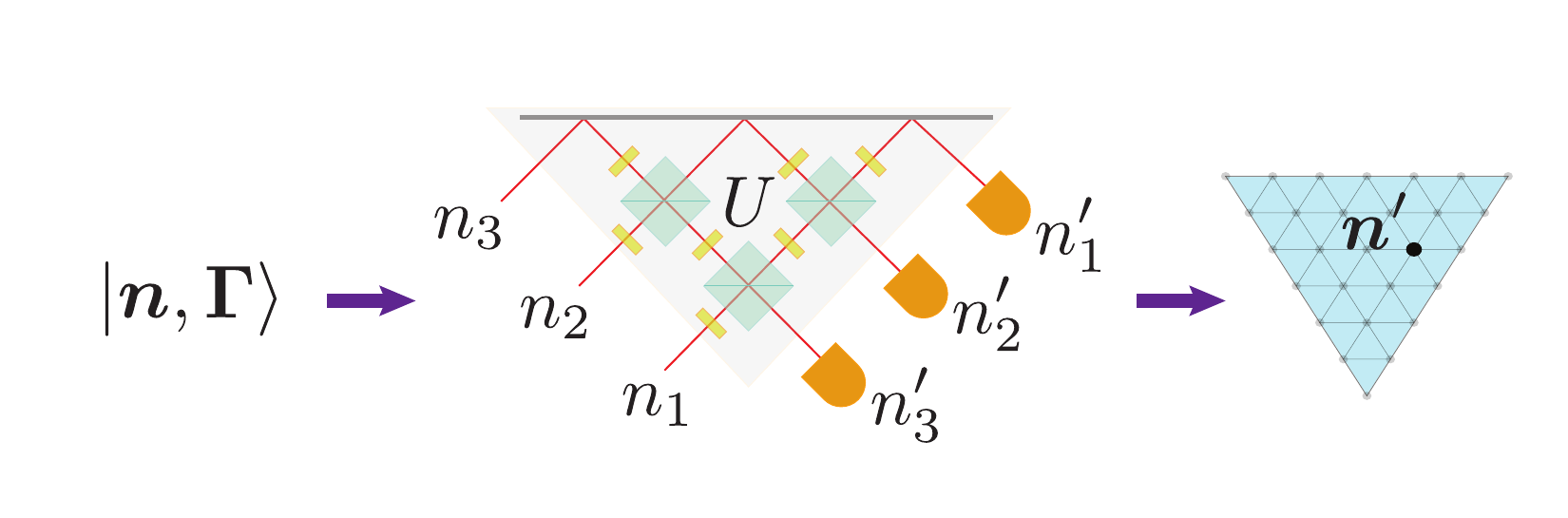}
\caption{Schematic of our setup for multiport, multiphoton interferometry, illustrated with $M=3$ ports. $N$ photons are prepared in a correlated mode state $\ket{\vect{n}, \vect{\Gamma}}$,  where the initial occupations $\vect{n} = (n_1, \ldots, n_M)$ describes the number of photons entering each port, and the matrix $\vect{\Gamma}$ the relevant parameters of the internal modes. The interferometer (shaded region) is described by a unitary $U$. At the output, photodetectors count the number of exiting photons at each port (without resolving internal modes), and the counts $n_i'$ determine the final occupations $\vect{n}'$, which can be viewed as points on a (final) occupation simplex (see Fig.~\ref{fig:occsimplex}). The multiphotonic scattering probability $P(\vect{n}'|U, \vect{n},\vect{\Gamma})$ gives the probability of  obtaining final occupations $\vect{n}'$ given the input state $\ket{\vect{n}, \vect{\Gamma}}$ and the unitary $U$. }
\label{fig:interferometer}
\end{figure}

\section{Setup}
\label{sec:setup}

\subsection{Interferometer and scattering probability}

The general setting for our discussion is an $M$-port linear 
passive interferometer in which $N$ incoming 
photons are distributed among the input ports $i=1,\ldots,M$ 
according to a distribution of input occupations $\vect{n}=
(n_1,n_2,\ldots,n_M)$, as illustrated in 
Fig.~\ref{fig:interferometer}.  At the single-photon 
level, the interferometer functions as a unitary 
``scattering'' matrix $U\in U(M)$ acting on an $M$-dimensional ``port'' Hilbert space 
$\mathcal{H}_p\cong\mathbb{C}^M$ with canonical basis 
$\{\ket{i}\}$, where $i$ labels the port. The interferometer 
scatters each photon from an incoming port $\ket{i}$ to all 
other ports according to $\ket{i}\to\sum_k\ket{k}U_{ki}$. At 
the output, photon-number-resolving detectors at each port 
register a count distribution (henceforth, \emph{output 
occupations}) $\vect{n}'=(n_1',n_2',\ldots,n_M')$, with 
$\sum_{i=1}^M n_i=\sum_{i=1}^M n_i'=N$ for ideal detectors. 

In addition to the port degree of freedom, single-photon 
states also come with other degrees of freedom such as 
polarization, frequency, etc.,  that, for the purpose of 
this paper, will be lumped into a single effective 
\emph{internal degree of freedom}.  
Without loss of generality, internal states are 
elements of a Hilbert space 
$\mathcal{H}_{\overline{p}} \cong\mathbb{C}^M$.
 The internal degree of 
freedom is left untouched by the  interferometer and is unresolved (traced out) by the output 
detectors. Correlations between the port and internal degrees of freedom are nonetheless central to 
multiphoton interference, since they determine the degree of \emph{distinguishability} among the photons --
how much ``which-path'' (i.e., ``port'') information can be inferred from the internal state. This information is encoded in an $M \times M$ Hermitian \emph{distinguishability matrix} $\vect{\Gamma}$ with diagonal entries equal to $1$. For the moment, it suffices to know that  the scenario of \emph{perfectly indistinguishable} photons is characterized by a distinguishability matrix with all elements $\Gamma_{i j} = 1$, and the case of \emph{perfectly distinguishable} photons corresponds to the case of the unit matrix: $\Gamma_{i j} = \delta_{ij}$.  The general case will be defined precisely later.

\begin{figure}[ht]
\centering
 \includegraphics[scale=0.28]{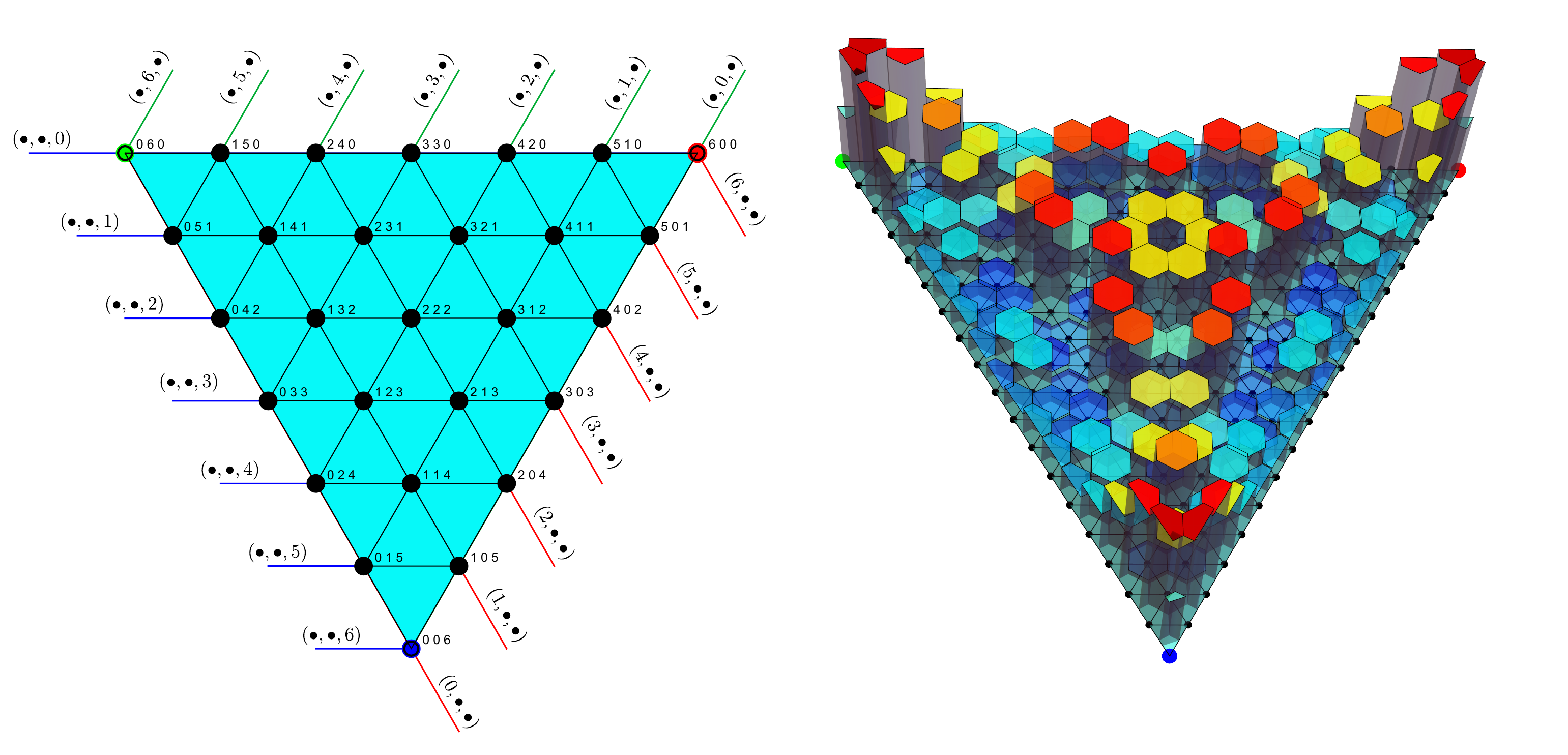}
\caption{Left: Occupation simplex for a three-port setting with $N=6$ photons. Points denote  possible occupation distributions $\vect{n}=(n_1,n_2,n_3)$ for a given $N$.  Points at the vertices of the simplex (colored) correspond to single-mode occupations.  
Right: Example of a multiphotonic scattering probability $P(\vect{n}'|U, \vect{n},\vect{\Gamma})$ for $N=18$ indistinguishable photons (${\Gamma}_{ij} = {1}$) and a three-port ``tritter'' interferometer ($U_{jk}= e^{\ii \frac{2 \pi}{3}j k }/\sqrt{3}$),  plotted on the simplex of final occupations $\vect{n}'$. The initial occupations are $\vect{n}= (3,5,10)$.  }
\label{fig:occsimplex}
\end{figure}

With these definitions, the main object of interest 
is the \emph{multiphotonic scattering probability} 
(MSP) $P(\vect{n}'|U,\vect{n},\vect{\Gamma})$  of 
obtaining output occupations $\vect{n}'$  given 
three conditioning parameters: the scattering matrix 
$U$, the distribution $\vect{n}$ of input 
occupations, and the distinguishability matrix  
$\vect{\Gamma}$. For a fixed $N$, the initial and 
final occupations $\vect{n}$ and $\vect{n}'$  can be 
visualized as points on an \emph{occupation simplex}---the integer points  in the convex hull of the $M$ points $(N,0,\ldots,0),\ (0,N,0,\ldots,0),\ \ldots,\ (0,0,\ldots,0,N)$ in $\mathbb{R}^{M}$, as  illustrated in Fig. \ref{fig:occsimplex}  for the case $N=6, M=3$. Viewed as a function of $\vect{n}'$ 
with fixed conditioning parameters, the MSP 
$P(\vect{n}'|U,\vect{n},\vect{\Gamma})$ can be 
visualized as a probability distribution on the 
occupation simplex, as illustrated in Fig. 
\ref{fig:occsimplex} for $N=18, M=3$.  

Our objective is to understand the behavior of the MSP $P(\vect{n}'|U,\vect{n},\vect{\Gamma})$ in the \emph{asymptotic limit}, which will be understood to be the limit $N \rightarrow \infty$  while holding  the \emph{normalized} occupations $\ov{n}= \vect{n}/N$ and $\ov{n}'= \vect{n}'/N$ fixed (or more precisely, with sequences of occupations $\vect{n}$ and $\vect{n}'$ such that $\vect{n}/N$ and $\vect{n}'/N$ converge to $\ov{n}$ and $\ov{n}'$). The normalized occupations  $\ov{n}$ and $\ov{n}'$ can be regarded as  points of the standard probability simplex, here denoted by $\Delta$, with minimal spacing $\sim 1/N$, so that as $N \rightarrow \infty$, the discrete distribution (or more precisely, its coarse-grained behavior) is naturally compared to a smooth probability density on $\Delta$.

\subsection{Perfectly indistinguishable photons}

In the case of perfect indistinguishability ($\Gamma_{ij}=1$), all 
photons are described by a single-particle state of the form 
$\ket{i}_p \ket{\chi}_{\overline{p}}$, where the internal state 
$\ket{\chi}_{\overline{p}}$ is common to \emph{all} $N$ incoming 
photons. Since this internal state is uncorrelated with the ports 
and unaffected by the interferometer, the effective single-particle 
Hilbert space is the port space $\mathcal{H}_p$. An $N$-photon 
state is then a state in the $N$-particle bosonic Fock space 
$\mathcal{F}^{(N)} \cong \Sym_N(\mathbb{C}^M)$ associated with the 
single-particle space $\mathcal{H}^{(1)}= 
\mathcal{H}_p=\mathbb{C}^M$, where $\Sym_N(V)$ is the symmetric 
subspace of $V^{\otimes N}$. The initial and final port occupations 
$\vect{n}$ and $\vect{n}'$ correspond to Fock states
\begin{equation}
\label{eq:indistfock}
    \ket{\vect{n}} = \prod_{i=1}^{M} \frac{ (a_i^\dagger)^{n_i}}{\sqrt{n_i!}}\ket{\vect{0}}, \qquad \ket{\vect{n}'} = \prod_{i=1}^{M} \frac{ (a_i^\dagger)^{n_i'}}{\sqrt{n_i'!}}\ket{\vect{0}},
\end{equation}
and the MSP is the squared modulus of the transition amplitude 
$\bra{\vect{n}'} D^{(N)}(U) \ket{\vect{n}}$, where $D^{(N)}(U)$ 
represents the single-particle transformation 
$\ket{i}\to\sum_k\ket{k}U_{ki}$ in $\mathcal{F}^{(N)}$, and 
transforms creation operators as $a_i^\dagger \rightarrow 
b_i^\dagger$, with
\begin{align}
\label{eq:abdaggerindist}
 b_i^\dagger = D^{(N)}(U) a_i^\dagger D^{(N)}(U^\dagger) = \sum_{k} a_k^\dagger U_{ki}\, .
\end{align}
Omitting  the distinguishability matrix, which is trivial in this case, the MSP for perfectly indistinguishable photons is
 \begin{align}
 \label{eq:MSPindist}
P(\vect{n}'| U, \vect{n} ) = \frac{1}{\prod_{i,j}n_i'!n_j!} \Big|\bra{\vect{0}}
\prod_{i,j}a_i^{n_i'}(b_j^\dagger)^{n_j} \ket{\vect{0}}\Big|^2. 
\end{align}
This transition amplitude has useful interpretations elsewhere: in group representation theory, it is a matrix element of the $U(M)$ Wigner $D$-matrix, or \emph{group function}, for $U$ in the symmetric representation; combinatorially, it equals the matrix permanent of an $N\times N$ augmentation of $U$ obtained by repeating columns and rows according to the initial and final occupation numbers. \cite{ChinGeneralizedconcurrence2018}\cite{shchesnovichASYMPTOTICEVALUATIONBOSONIC2013}
  
\subsection{Partial indistinguishability with correlated mode states}

Partial indistinguishability sets in when the internal degrees of freedom of the initial multiphotonic state are correlated with the port degrees of freedom. Multiphotonic states then belong to the bosonic $N$-particle Fock space $\mathcal{F}^{(N)} = \mathrm{Sym}_N(\mathcal{H}^{(1)})$ built from the composite single-particle Hilbert space $\mathcal{H}^{(1)} = \mathcal{H}_p \otimes \mathcal{H}_{\overline{p}}$.
Throughout this paper we restrict our attention to a family of initial multiphotonic states, which we term \emph{correlated mode states}. In these states, the $n_i$ photons entering port $i$ share the same single-particle state $\ket{i}\ket{\chi_i} \in \mathcal{H}_p \otimes \mathcal{H}_{\overline{p}}$, with normalized internal mode states $\ket{\chi_i}$ generally satisfying $0\le|\amp{\chi_i}{\chi_j}|\le1$ for $i \neq j$. Correlated mode states are then Fock states with respect to the orthogonal single-particle modes $\{ \ket{i}\ket{\chi_i} \}$, which span an $M$-dimensional subspace of the $M^2$-dimensional single-particle Hilbert space. Specifically, given orthonormal bases $\{ \ket{i} \}$ and $\{ \ket{\gamma} \}$ for $\mathcal{H}_p$ and $\mathcal{H}_{\overline{p}}$, let $a_{i, \gamma}^\dagger$ be creation operators for the product mode basis $\ket{i}\ket{\gamma}$. Photons in the single-particle mode $\ket{i}\ket{\chi_i}$ are then created by
\begin{equation}
\label{eq:aichi}
    a^\dagger_{i,\chi_i}=\sum_\gamma\amp{\gamma}{\chi_i}\,a_{i, \gamma}^\dagger,
\end{equation}
where $\amp{\gamma}{\chi_i}$ are the expansion coefficients of $\ket{\chi_i}$ in the basis for $\mathcal{H}_{\overline{p}}$. Because the detectors are blind to the internal modes, the only data relevant to the MSP are the overlaps between the internal states of the input photons. This information is encoded in the distinguishability matrix $\vect{\Gamma}$, the $M \times M$ Hermitian Gram matrix with elements
\begin{equation}
\label{eq:gram}
\Gamma_{ij}=\amp{\chi_j}{\chi_i}.
\end{equation}
(note the index placement relative to $\amp{\chi_j}{\chi_i}$). A correlated mode state, denoted $\ket{\vect{n}, \vect{\Gamma}}$, is any Fock state
\begin{equation}
\label{eq:cms}
\ket{\vect{n},\vect{\Gamma}}=\prod_{i=1}^{M}\frac{\big(a^\dagger_{i,\chi_i}\big)^{n_i}}{\sqrt{n_i!}}\ket{\vect{0}},
\end{equation}
with occupations $\vect{n}$ in a set of modes $\{ \ket{i}\ket{\chi_i} \}$ satisfying $\amp{\chi_j}{\chi_i}=\Gamma_{ij}$.

The family of correlated mode states is bracketed by the two limiting cases discussed earlier: perfectly indistinguishable photons, where all $\ket{\chi_i}$ coincide and $\Gamma_{ij}=1$ for all $i,j$; and perfectly distinguishable photons, where the $\ket{\chi_i}$ are mutually orthogonal and $\vect{\Gamma}=\one$. Intermediate cases of $\vect{\Gamma}$---for example, photons with partially overlapping polarizations---interpolate between these extremes.

Despite the formal similarity between the correlated-mode-state definition \eqref{eq:cms} and the Fock states $\ket{\vect{n}}$ of Eq.~\eqref{eq:indistfock}, the non-orthogonality of the $\ket{\chi_i}$ and the fact that the internal degree of freedom is not resolved by the detector make the MSP far less amenable to an expression like Eq.~\eqref{eq:MSPindist} in terms of Fock-space creation and annihilation operators. An exact expression is
\begin{align}
 \label{eq:MSPdist}
P(\vect{n}'|U,\vect{n},\vect{\Gamma})
=\frac{1}{\prod_i n_i!}
\sum_{\{n_{j,\gamma}'\}}^*
\frac{1}{\prod_{j,\gamma}n_{j,\gamma}'!}
\left|\bra{\vect{0}}
\left(\prod_{j,\gamma}
a_{j,\gamma}^{n_{j,\gamma}'}\right)
\left(\prod_i(b_{i,\chi_i}^\dagger)^{n_i}\right)
\ket{\vect{0}}\right|^2, 
\end{align}
where the asterisk denotes a constrained sum over joint output occupations $n_{j,\gamma}'$ of the product modes $\ket{i}\ket{\gamma}$ such that $n_j'=\sum_\gamma n_{j,\gamma}'$ for every output port $j$, and
\begin{equation}
    b^\dagger_{i,\chi_i}=\sum_{\gamma}\sum_{k} \amp{\gamma}{\chi_i}U_{ki}\,a_{k, \gamma}^\dagger,  
\end{equation}
in analogy to Eq.~\eqref{eq:abdaggerindist}.  As before, the terms in the sum can be rewritten in terms of $U(M^2)$ group functions or matrix permanents \cite{tichySamplingPartiallyDistinguishable2015}\cite{spivakGeneralizedInterferenceFermions2022}.

While \eqref{eq:MSPindist} and \eqref{eq:MSPdist} are exact Fock-space expressions for the MSP in the perfectly indistinguishable and partially indistinguishable cases, they are opaque and poorly suited to asymptotic analysis. In the next section we develop an alternative representation of the correlated mode state $\ket{\vect{n},\vect{\Gamma}}$ based on tensor powers of single-particle states. The resulting toroidal tensor power expansion provides a mathematical framework---and a physically intuitive picture---well suited to the asymptotics analysis of the MSP.

\section{Virtual states and the toroidal tensor power expansion}\label{sec:vs}

\subsection{Tensor power states and tensor power expansions}\label{sec:tpstpe}

For $N$ identical particles, each described by a Hilbert space $\mathcal{H}^{(1)}$, a tensor-power state $\ket{\psi^{\otimes N}} \in (\mathcal{H}^{(1)})^{\otimes N}$ is one in which all particles occupy the same single-particle state $\ket{\psi}\in\mathcal{H}^{(1)}$, and is thus represented as the tensor power $\ket{\psi}^{\otimes N}$. Since this state is invariant under permutations of the copies of $\mathcal{H}^{(1)}$, it lives in the symmetric subspace of $(\mathcal{H}^{(1)})^{\otimes N}$, the $N$-particle bosonic Fock space $\mathcal{F}^{(N)}=\mathrm{Sym}_N(\mathcal{H}^{(1)})$. Thus a tensor power state is a Fock state
\begin{equation}
\label{eq:tps}
    \ket{\psi^{\otimes N}}=\frac{(a^\dagger_\psi)^N}{\sqrt{N!}}\ket{\vect{0}}, \qquad a^\dagger_\psi=\sum_{j}\psi_{j}\,a^\dagger_{j},
\end{equation}
where $a_\psi^\dagger$ creates a particle in the mode $\ket{\psi}$, a linear combination of single-particle modes with respect to a canonical basis $\{ \ket{j} \}$ for $\mathcal{H}^{(1)}$, i.e., $\ket{\psi} = \sum_{j}\psi_j \ket{j}$. Tensor power states are also known as \emph{i.i.d.\ states} or \emph{$SU(d)$-coherent states}.

The set of tensor power states spans $\mathcal{F}^{(N)}$ overcompletely, so any bosonic multiphoton state $\ket{\Psi} \in \mathcal{F}^{(N)}$ can be expanded, typically in many ways, as a superposition of tensor powers
\begin{equation}
\label{eq:tpe}
\ket{\Psi}=\sum_{s=1}^{s_{\max}}c_s\ket{\psi_s^{\otimes N}},
\end{equation}
(or its continuous generalization) where the $\ket{\psi_s}$ are generally not mutually orthogonal (so $\sum_s |c_s|^2 \neq 1$ in general). The minimal number of terms in any such expansion is, however, an intrinsic property of the state---its \emph{symmetric rank} $\mathrm{R}_S(\ket{\Psi})$~\cite{comonSymmetricTensorsSymmetric2008}. An $SU(d)$
coherent state \cite{perelomov1972}\cite{Zhandcoherentstates1090}
has $\mathrm{R}_S=1$, and so does a Fock state with a single occupied input mode, but Fock states with multi-mode occupations, such as correlated mode states $\ket{\vect{n}, \vect{\Gamma}}$, have $\mathrm{R}_S>1$ \cite{carliniSolutionWaringProblem2012,buczynskaWaringDecompositionsMonomials2013}. 

The symmetric rank provides a convenient characterization of multiphotonic interference. To see this, let $\Pi_{\vect{n}}$ be the projector onto the subspace of Fock space with port occupations $\vect{n}$, so that the probability of finding occupations $\vect{n}$ in the state $\ket{\Psi}$ is $P(\vect{n}|\Psi)=\bra{\Psi}\Pi_{\vect{n}}\ket{\Psi}$. If $\ket{\Psi}$ has $R_S=1$, then $\ket{\Psi}=\ket{\psi^{\otimes N}}$ for some single-particle state $\ket{\psi}$, in which case $P(\vect{n}|\psi^{\otimes N})$ is a multinomial distribution. Denoting such distributions by $\Mult(\vect{n}|\vect{p})$, with
\begin{equation}
\label{eq:multdist}
\Mult(\vect{n}|\vect{p}):=\frac{N!}{\prod_{i=1}^M n_i!}\prod_{i=1}^M p_i^{\,n_i},
\end{equation}
we have $P(\vect{n}|\psi^{\otimes N})=\Mult(\vect{n}|\vect{p})$, where the sampling probabilities are $p_k=\bra{k}\rho_\psi\ket{k}$ with $\rho_\psi=\Tr_{\bar p}\ket{\psi}\bra{\psi}$. If instead $\ket{\Psi}$ has symmetric rank $R_S>1$, the expectation value $\bra{\Psi}\Pi_{\vect{n}}\ket{\Psi}$ necessarily exhibits interference between the distinct tensor-power terms in \eqref{eq:tpe}, and $P(\vect{n}|\Psi)$ splits into a mixture of multinomials plus an interference term present only when $R_S>1$,
\begin{equation}
\label{eq:tpeinterf}
P(\vect{n}|\Psi)=\underbrace{\sum_s|c_s|^2\,P(\vect{n}|\psi_s^{\otimes N})}_{\text{multinomial mixture}}
\;+\;\underbrace{\sum_{s\ne s'}c_s c_{s'}^*\,\bra{\psi_{s'}^{\otimes N}}\Pi_{\vect{n}}\ket{\psi_s^{\otimes N}}}_{\text{interference}}.
\end{equation}
Since $\sum_s|c_s|^2\neq1$ in general, this interpretation is only approximate; in the exact case the sums over $\vect{n}$ of the two terms would equal one and zero, respectively. Nevertheless, since $|\amp{\psi_s^{\otimes N}}{\psi^{\otimes N}_{s'}}| = |\amp{\psi_s}{\psi_{s'}}|^N \rightarrow 0$ as $N \rightarrow \infty$ for $s\neq s'$, the interpretation is expected to sharpen asymptotically.

Tensor power expansions also make it simple to track the evolution of a multiphotonic state under a single-particle operation---a feature we will exploit repeatedly in what follows. For any invertible $g\in GL(M,\mathbb{C})$, the Fock-space evolution $D^{(N)}(g)$ acts on the expansion as
\begin{equation}
\label{eq:tpe_evol}
\ket{\Psi}=\sum_{s=1}^{s_{\max}}c_s\ket{\psi_s^{\otimes N}} \longrightarrow D^{(N)}(g)\ket{\Psi} =\sum_{s=1}^{s_{\max}}c_s\ket{(g \psi_s)^{\otimes N}};
\end{equation}
i.e., the single-particle states of the expansion are simply replaced by their images under $g$. Since $g$ is invertible, both expansions involve the same number of terms, and as a consequence, the symmetric rank is invariant under invertible single-particle transformations.

\subsection{Virtual states and their tensor powers}\label{sec:virtualstates}

For the tensor power expansion of multiphoton states in the setup of Sec.~\ref{sec:setup}, it is convenient to introduce a parameterized family of single-particle states $\ket{\vect{\nu},\vect{\Gamma}} \in \mathcal{H}_p \otimes \mathcal{H}_{\overline{p}}$, which we refer to as \emph{virtual states}. A virtual state is parameterized by an $M$-component probability vector $\vect{\nu} = (\nu_1, \nu_2, \ldots, \nu_M)$ --- the \emph{virtual weights} --- and an $M\times M$ Gram matrix $\vect{\Gamma}$ --- the \emph{distinguishability matrix} --- associated with a set of normalized internal mode states $\{\ket{\chi_1}, \ket{\chi_2}, \ldots, \ket{\chi_M}\}$, according to
\begin{equation}
\label{eq:vs}
\ket{\vect{\nu},\vect{\Gamma}}=\sum_{i=1}^{M}\sqrt{\nu_i}\,\ket{i}\ket{\chi_i}, \qquad \Gamma_{ij} = \amp{\chi_j}{\chi_i},
\end{equation}
with $\Gamma_{ii}=1$. Since the $\ket{\chi_i}$ are not uniquely specified by $\vect{\Gamma}$, the set $\{\ket{\chi_i}\}$ can be thought of as a representative of the equivalence class of states with Gram matrix $\vect{\Gamma}$. As  the internal modes are not resolved in detection, only the overlaps $\amp{\chi_j}{\chi_i}$ matter for the MSP, and any class representative suffices.

From its definition, a virtual state defines a port-space marginal density matrix $\mathlarger\margvs=\Tr_{\bar p}\,\ket{\vect{\nu},\vect{\Gamma}}\bra{\vect{\nu},\vect{\Gamma}}$, or \emph{marginal virtual state}, with elements
\begin{equation}
\label{eq:rho}
\bra{i}\margvs\ket{j}=\sqrt{\nu_i\nu_j}\,\Gamma_{ij}.
\end{equation}
Since $\Gamma_{ii}=1$, the diagonal elements of $\mathlarger\margvs$ are the virtual weights $\nu_i$, the probability of finding the particle in port $i$. From Eq.~\eqref{eq:rho}, the parameters $(\vect{\nu},\vect{\Gamma})$ can be recovered from the marginal virtual state. 

Passing to the multiphoton setting, we denote by $\ket{ (\vect{\nu}, \vect{\Gamma})^{\otimes N}}$ the tensor power state, or $SU(M^2)$-coherent state, in which all $N$ photons are in the same single-particle state $\ket{ \vect{\nu}, \vect{\Gamma}}$. The probability of finding the $N$ photons distributed among the ports according to occupations $\vect{n}$ is then the multinomial probability \eqref{eq:multdist} with sampling probability vector $\vect{p}= \vect{\nu}$:
\begin{equation}
\label{eq:vsoccprob}
 \bra{ (\vect{\nu}, \vect{\Gamma})^{\otimes N}}\Pi_{\vect{n} } \ket{ (\vect{\nu}, \vect{\Gamma})^{\otimes N}} = \Mult(\vect{n}|\vect{\nu}).
\end{equation}
Suppose now that $\vect{\nu}$ is arbitrary except that all $\nu_i \neq 0$.  Then, as shown in ~\ref{TPprojappendix}, the projection $\Pi_{\vect{n} }\ket{ (\vect{\nu}, \vect{\Gamma})^{\otimes N}}$ onto the subspace of Fock space with occupations $\vect{n}$ is proportional to the correlated mode state $\ket{\vect{n},\vect{\Gamma}}$ with the same distinguishability matrix $\vect{\Gamma}$, where the proportionality constant follows from \eqref{eq:vsoccprob}:
\begin{equation}
\label{eq:proj}
\ket{\vect{n};\vect{\Gamma}}=\frac{1}{\sqrt{\Mult(\vect{n}|\vect{\nu})}}\,
\Pi_{\vect{n}}\ket{(\vect{\nu},\vect{\Gamma})^{\otimes N}}.
\end{equation}
This relation is the starting point of the toroidal tensor power expansion.
An important aspect of this expansion is the choice of the parameter $\vect{\nu}$. While   \eqref{eq:proj} is valid for any virtual weight $\vect{\nu}$, the choice $\vect{\nu} = \ov{n}$ is optimal for the asymptotic analysis, as it is the \emph{only} choice ensuring that the normalization factor $1/\sqrt{\Mult(\vect{n}|\vect{\nu})}$ in \eqref{eq:proj} is controlled polynomially rather than exponentially. Indeed, by Sanov's theorem \cite{coverElementsInformationTheory2005,csiszarSimpleProofSanovs2006}, the multinomial distribution obeys
\begin{equation}
\label{eq:typebound}
(N+1)^{-M}e^{-N D(\ov{n}\|\vect{p})}\le\Mult(\vect{n}|\vect{p})\le e^{-N D(\ov{n}\|\vect{p})},
\end{equation}
where $D(\vect{q}\|\vect{p})=\sum_i q_i\ln(q_i/p_i)$ is the Kullback--Leibler relative entropy, which vanishes only when $\ov{n} =\vect{p}$ and is otherwise strictly positive. Thus $\Mult(\vect{n}|\vect{p})$ decays exponentially in $N$ unless the sampling distribution matches the normalized occupations, in which case it is only polynomially bounded. This fact underlies all the asymptotic results in this paper.

\subsection{Toroidal tensor power expansion of correlated mode states}\label{ttpecms}

 The toroidal expansion is a special type of continuous tensor power expansion of the correlated mode state $\ket{\vect{n}, \vect{\Gamma} }$ in \eqref{eq:cms}, which by \eqref{eq:tpe_evol} induces a corresponding expansion of the final state $D^{(N)}(U \otimes \mathbf{1})
\ket{\vect{n},\vect{\Gamma}}$ that results from scattering each photon through the interferometer. In either case, the single-particle states of the expansion are virtual states whose parameters $(\vect{\nu}, \vect{\Gamma})$ are functions of the $(M-1)$-dimensional torus $T$ defined by the abelian group of (relative) phase-twisting $M \times M$ matrices
\begin{equation}
    t_{\vect{\phi}} =  \sum_{k}e^{\ii\phi_k}\ket{k}\bra{k}, \qquad \phi_1 :=0,  
\end{equation}
where $\vect{\phi} = (\phi_1, \phi_2, \ldots \phi_M)$ are the angular coordinates of $T$, with $\phi_1 = 0$.

For the toroidal expansion of $\ket{\vect{n},\vect{\Gamma}}$, the $T$-dependent virtual state parameters, denoted $(\vect{\nu}_{\vect{\phi}}, \vect{\Gamma}_{\vect{\phi}})$, are as follows: the virtual weights $\vect{\nu}_{\vect{\phi}}$ are taken independent of $\vect{\phi}$ and, crucially, \emph{set to the values of the reduced initial occupations $\ov{n}= \vect{n}/N$}---the asymptotically optimal choice explained previously; the distinguishability matrix $\vect{\Gamma}_{\vect{\phi}}$ is taken to be a ``phase-twisted'' version of $\vect{\Gamma}$, with elements
$
    (\Gamma_{\vect{\phi}})_{i j} = \Gamma_{i j}e^{\ii (\phi_i-\phi_j)}. 
$
The corresponding virtual states can be expressed in terms of the internal-mode states $\{ \ket{\chi_i } \}$ of the initial multiphotonic state as
\begin{equation}
\label{eq:vsrot}
\ket{\vect{\nu}_{\vect{\phi}},\vect{\Gamma}_{\vect{\phi}}}= \sum_{i=1}^{M}\sqrt{\overline{n}_i}\,\ket{i}\ket{\chi_i}e^{\ii \phi_i}. 
\end{equation}
All initial virtual states thus lie on the same orbit under the phase-twisting group acting on the port side, i.e., $\ket{\vect{\nu}_{\vect{\phi}},\vect{\Gamma}_{\vect{\phi}}}=\left(t_{\vect{\phi}-\vect{\varphi}}\otimes\mathbf{1}\right)
\ket{\vect{\nu}_{\vect{\varphi}},\vect{\Gamma}_{\vect{\varphi}}}$. With these notions now in place, we have:
\medskip
\begin{proposition}
\label{prop:toroidalexpansion}
(Toroidal expansion) The toroidal expansion of the correlated mode state $\ket{\vect{n},\vect{\Gamma}}$ is
\begin{equation}
\label{eq:contdec}
\ket{\vect{n},\vect{\Gamma}}=\frac{1}{\sqrt{\Mult(\vect{n}|\ov{n})}}\int_T d\mu(\vect\phi)\,
e^{-\ii\vect{n}\cdot\vect\phi}\,\ket{(\vect{\nu}_{\vect{\phi}},\vect{\Gamma}_{\vect\phi})^{\otimes N}}, \qquad (\vect{\nu}_{\vect\phi},  \vect{\Gamma}_{\vect\phi}) =  (\ov{n}, t_{\vect{\phi}} \vect{\Gamma} t^\dagger_{\vect{\phi}}),
\end{equation}
where $d\mu$ is the uniform (Haar) measure on $T$, normalized so that $\int_T d\mu(\vect\phi)=1$.
\end{proposition}

The toroidal expansion follows from \eqref{eq:proj} by choosing $\vect{\nu} = \ov{n}$ and replacing the projector $\Pi_{\vect{n}}$ with its Fourier expansion, which can be conveniently written in terms of the phase-twist matrix as
\begin{equation}
\label{eq:projint}
\Pi_{\vect{n}}=\int_T d\mu(\vect\phi)\,e^{-\ii\vect{n}\cdot\vect\phi}\,(t_{\vect\phi} \otimes \mathbf{1})^{\otimes N}.
\end{equation}
Using $\ket{\vect{\nu}_{\vect{\phi}},\vect{\Gamma}_{\vect{\phi}}}=\left(t_{\vect{\phi}}\otimes\mathbf{1}\right)\ket{\ov{n},\vect{\Gamma}}$, one obtains \eqref{eq:contdec}. This expansion is over a continuous family of tensor power states and so is not minimal in the sense of attaining the symmetric rank of the state; nevertheless, it is the limiting form, as $N \rightarrow \infty$, of the minimal tensor power expansion realizing the symmetric rank, found in \cite{carliniSolutionWaringProblem2012,buczynskaWaringDecompositionsMonomials2013}.

\subsection{Toroidal Expansion of the scattered multiphotonic  state}

With the correlated mode state $\ket{\vect{n},\vect{\Gamma}}$ expanded as a superposition of tensor powers of the phase-twisted virtual states $\ket{\vect{\nu}_{\vect\phi},\vect{\Gamma}_{\vect\phi}}$, the scattering by $U$ can be tracked at the level of individual particles. By \eqref{eq:tpe_evol}, the post-scattering multiphotonic state is again a toroidal tensor power expansion, but over the \emph{scattered} virtual states obtained by applying $U\otimes\one$ to the $\ket{\vect{\nu}_{\vect\phi},\vect{\Gamma}_{\vect\phi}}$. Denoting the parameters of this family by $(\widetilde{\vect{\nu}}_{\vect{\phi}},\widetilde{\vect{\Gamma}}_{\vect{\phi}})$, the toroidal expansion of the scattered multiphotonic state is
\begin{equation}
\label{eq:contdecfinal}
    D^{(N)}(U \otimes \mathbf{1}) \ket{\vect{n}, \vect{\Gamma}} = \frac{1}{\sqrt{\Mult(\vect{n}|\ov{n})}}\int_T d\mu(\vect\phi)\,
e^{-\ii\vect{n}\cdot\vect\phi}\,\ket{( \widetilde{\vect{\nu}}_{\vect{\phi}},\widetilde{\vect{\Gamma}}_{\vect{\phi}})^{\otimes N}},
\end{equation}
where
\begin{equation}
\label{eq:defscatvs}
\ket{\widetilde{\vect{\nu}}_{\vect{\phi}},\widetilde{\vect{\Gamma}}_{\vect{\phi}}  } := U\otimes\one\ket{\vect{\nu}_{\vect\phi},\vect{\Gamma}_{\vect\phi}}.
\end{equation}
Thus, the scattered multiphotonic state can be interpreted as a coherent superposition of $N$-photon coherent-state scattering ``experiments'', each parametrized by a phase vector $\vect\phi$, and which in isolation would give an MSP equal to the multinomial probability $\Mult(\vect{n}'|\widetilde{\vect{\nu}}_{\vect{\phi}})$.

\begin{figure}[t]
\centering
 \includegraphics[scale=0.45]{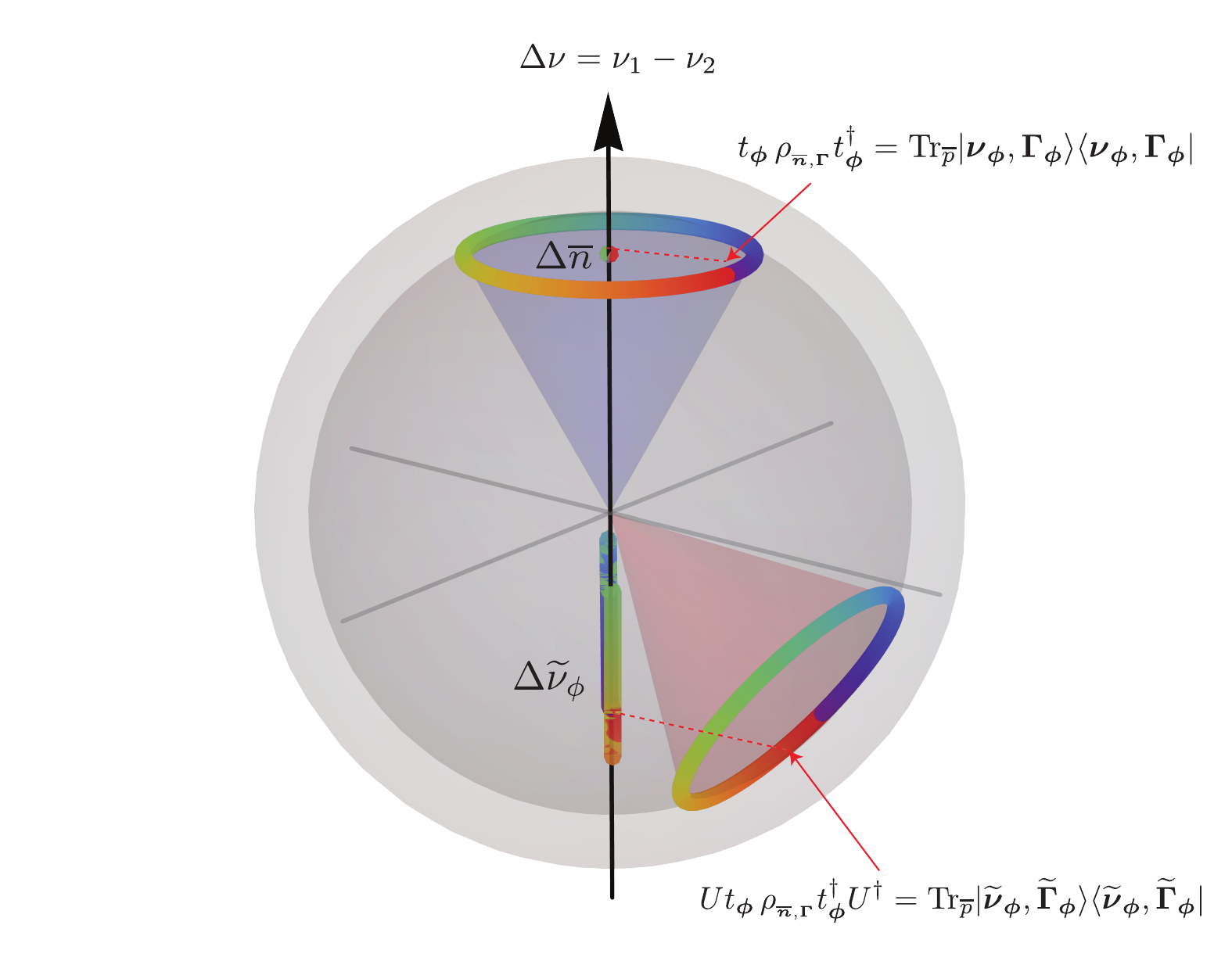}
\caption{Graphical illustration of the scattering of virtual states for the case $M=2$, in terms of marginal virtual states represented as points in the interior of the Bloch sphere.  The circles in the two cones represent  the marginal states for the  virtual states $\ket{\vect{\nu}_{\vect\phi},\vect{\Gamma}_{\vect\phi}}$ and $\ket{\widetilde{\vect{\nu}}_{\vect{\phi}},\widetilde{\vect{\Gamma}}_{\vect{\phi}}  }$ in the toroidal expansions of $\ket{\vect{n}, \vect{\Gamma}}$  and $D^{(N)}(U\otimes \one)\ket{\vect{n}, \vect{\Gamma}}$ respectively.  The $z$ axis plays the role of the $1$-simplex with projections onto the $z$ axis representing  the difference $\Delta \nu = \nu_1-\nu_2$ in the virtual weights of the virtual state. All initial states project to the same point $\Delta \overline{n} = \overline{n}_1 - \overline{n}_2$ on the $z$ axis. The scattered states project onto different points $\Delta\widetilde{\nu} = \widetilde{\nu}_1 - \widetilde{\nu}_2$, depending on the values of $\phi$. The density of these projected points  is the classical measure. }
\label{fig:cones}
\end{figure}

In contrast to the parameters $(\vect{\nu}_{\vect{\phi}}, \vect{\Gamma}_{\vect{\phi}})$ of the initial virtual states, the scattered-state parameters $(\widetilde{\vect{\nu}}_{\vect{\phi}},\widetilde{\vect{\Gamma}}_{\vect{\phi}} )$ depend on $\vect{\phi}$, and on the other parameters of the problem, in a more complicated way, through functional relations
$
\widetilde{\vect{\nu}}_{\vect{\phi}}= F_U( \vect{\nu}_{\vect{\phi}}, \vect{\Gamma}_{\vect{\phi}})
$,
$\widetilde{\vect{\Gamma}}_{\vect{\phi}} = G_U( \vect{\nu}_{\vect{\phi}}, \vect{\Gamma}_{\vect{\phi}}),
$
depending on the scattering matrix $U$. These relations follow from \eqref{eq:defscatvs} by expanding the right-hand side in the standard virtual-state form \eqref{eq:vs}, with new normalized internal-mode states $\ket{\tilde{\chi}_{\vect{\phi},i}}$:
\begin{equation}
\label{eq:scatvsrot}
\ket{\widetilde{\vect{\nu}}_{\vect{\phi}},\widetilde{\vect{\Gamma}}_{\vect{\phi}}  } = \sum_i \sqrt{ \widetilde{\nu}_{\vect{\phi}}{}_{,i} }\ket{i} \ket{ \widetilde{\chi}_{\vect{\phi}}{}_{,i}}, 
\end{equation}
and identifying $\widetilde{ \vect{\Gamma} }_{\vect{\phi}}$ as the Gram matrix with elements $\amp{\widetilde{\chi}_{\vect{\phi}}{}_{,j}}{\widetilde{\chi}_{\vect{\phi}}{}_{,i}}$.
Letting $[\vect{\nu}]$ denote the diagonal matrix with entries $\vect{\nu}$, the scattered-state parameters are
\begin{equation}
\label{eq:nutilde}
\widetilde{\vect{\nu}}_{\vect{\phi}}  = \mathrm{Diag} \left(  U\, [ {\vect{\nu}_{\vect{\phi}}} ]^{1/2}\, \vect{\Gamma}_{\vect{\phi}}\, [ {\vect{\nu}_{\vect{\phi}}} ]^{1/2}\, U^\dagger \right),
\end{equation}
where $\mathrm{Diag}(\ldots)$ denotes the diagonal elements of its argument, and
\begin{equation}
\widetilde{\vect{\Gamma}}_{\vect{\phi}} = [\widetilde{\vect{\nu}}_{\vect{\phi}}]^{-1/2}  U\, [ {\vect{\nu}_{\vect{\phi}}} ]^{1/2}\, \vect{\Gamma}_{\vect{\phi}}\, [ {\vect{\nu}_{\vect{\phi}}} ]^{1/2}\, U^\dagger \, [\widetilde{\vect{\nu}}_{\vect{\phi}}]^{-1/2}. 
\end{equation}
Of these two relations, \eqref{eq:nutilde} will be particularly important for the asymptotic analysis. For later convenience, we introduce the marginal virtual state
\begin{equation}
\label{eq:margn}
\margvsn := [\ov{n}]^{1/2}\, \vect{\Gamma}\,  [\ov{n}]^{1/2},
\end{equation}
the partial density matrix on the port space of the initial virtual state $\ket{\vect{\nu}_{\vect{\phi}}, \vect{\Gamma}_{\vect{\phi}}}$ at $\vect{\phi}=0$. In terms of $U$ and the parameters of $\ket{\vect{n}, \vect{\Gamma}}$, the scattered virtual weights are then
\begin{equation}
\label{eq:clasmapdef}
  \widetilde{\vect{\nu}}_{\vect{\phi}}  =  \mathrm{Diag} \left(  U\, t_{\vect{\phi}}\, \margvsn\,  t^\dagger_{\vect{\phi}}  U^\dagger \right).
\end{equation}
With all other parameters fixed, this relation defines a map---the ``classical map''---from $T$ to the probability simplex, central to the asymptotic analysis.
Figure \ref{fig:cones} illustrates the virtual state construction in terms of $\mathlarger \margvs$, and the resulting classical map for the case $M=2$.

\subsection{Toroidal expansion of the multiphotonic scattering probability}\label{teMSP}

Proposition~\ref{prop:toroidalexpansion} gives us the tools to compute the MSP when the incoming multiphotonic state is a correlated mode state: taking the expectation value of the projector $\Pi_{\vect{n}'}$ onto the final occupations in the scattered state $D^{(N)}(U \otimes \mathbf{1}) \ket{\vect{n}, \vect{\Gamma}}$, and expanding it using \eqref{eq:contdecfinal}, we obtain:
\medskip
\begin{proposition}
\label{prop:MSPdoubleintegral}
(Integral form of the MSP) The MSP can be written as a double integral on the torus 
\begin{equation}
\label{eq:probtorus}
P(\vect{n}'|U,\vect{n},\vect{\Gamma})=\frac{1}{\Mult(\vect{n}|\ov{n})}
\int_{T\times T} \! \! \! \! d\mu(\vect{\varphi})\,d\mu(\vect\phi)\, \, 
e^{\,\ii\vect{n}\cdot(\vect{\varphi}-\vect\phi)}\,
\, \bra{(\widetilde{\vect{\nu}}_{\vect{\varphi}},\widetilde{\vect{\Gamma}}_{\vect{\varphi}} )^{\otimes N}}\Pi_{\vect{n}'}\ket{(\widetilde{\vect{\nu}}_{\vect{\phi}},\widetilde{\vect{\Gamma}}_{\vect{\phi}} )^{\otimes N}} . 
\end{equation}
\end{proposition}
This integral can be thought of as a complex generalization of a multinomial mixture, since the integrand $\bra{(\widetilde{\vect{\nu}}_{\vect{\varphi}},\widetilde{\vect{\Gamma}}_{\vect{\varphi}} )^{\otimes N}}\Pi_{\vect{n}'}\ket{(\widetilde{\vect{\nu}}_{\vect{\phi}},\widetilde{\vect{\Gamma}}_{\vect{\phi}} )^{\otimes N}}$ can be written in a form similar to a multinomial distribution,
\begin{equation}
\bra{(\widetilde{\vect{\nu}}_{\vect{\varphi}},\widetilde{\vect{\Gamma}}_{\vect{\varphi}} )^{\otimes N}}\Pi_{\vect{n}'}\ket{(\widetilde{\vect{\nu}}_{\vect{\phi}},\widetilde{\vect{\Gamma}}_{\vect{\phi}} )^{\otimes N}} = \Mult(\vect{n}'|\vect{\xi}) = N! \prod_{k=1}^{M}   \frac{ {\xi}_k^{n_k'} }{n_k'!} ,  
\end{equation}
except the $\xi_k$ are no longer sampling probabilities but rather generally complex numbers
\begin{equation}
{\xi}_k =\bra{k} U t_{\vect{\phi}}\,  \margvsn\,  t_{\vect{\varphi}}^{\dagger} U^\dagger \ket{k},
\end{equation}
depending only on the single-particle marginal virtual state $\mathlarger \margvsn$. When $\vect{\phi} = \vect{\varphi}$, the $\xi_k$ coincide with the scattered virtual weights $\widetilde{\nu}_{\vect{\phi},k}$ and can be interpreted as genuine sampling probabilities.

Two points are worth highlighting concerning expansion \eqref{eq:probtorus}. First, since the integrand depends on $\vect{\Gamma}$ only through the marginal virtual state $\mathlarger \margvsn$, all the internal-mode information relevant to the MSP is contained in the distinguishability matrix, as stated previously. Second, viewing \eqref{eq:probtorus} in light of decomposition \eqref{eq:tpeinterf} identifies the source of multiphotonic interference: along the ``diagonal'' domain $\vect{\varphi} \simeq \vect\phi$, the integral produces a uniformly-weighted mixture of approximately multinomial distributions concentrated at occupations $\vect{n}' \simeq N \widetilde{\vect\nu}_{\vect\phi}$; the complementary ``off-diagonal'' domain contributes the interference terms, with phases coming from the exponential $e^{ \ii \vect{n}\cdot(\vect{\phi} - \vect{\varphi}) }$ and from the phases of the monomials $\prod_{k=1}^{M}   \xi_k^{n_k'}$.

\subsection{Relevant contributions to the MSP for large $N$}

The integrand of the toroidal expansion \eqref{eq:probtorus} is a complex function on $T \times T$ whose variations in magnitude and phase grow increasingly fast with $N$. Thus, at large $N$, it becomes possible to distinguish regions of $T \times T$ that contribute negligibly to the integral from those that potentially dominate the MSP $P(\vect{n}'|U,\vect{n}, \vect{\Gamma})$. A preliminary assessment of the relevance of a region around a given pair of angles $(\vect{\phi}, \vect{\varphi})$ comes from bounding the absolute value of the integrand,
$\left|\bra{(\widetilde{\vect{\nu}}_{\vect{\varphi}},\widetilde{\vect{\Gamma}}_{\vect{\varphi}} )^{\otimes N}}\Pi_{\vect{n}'}\ket{(\widetilde{\vect{\nu}}_{\vect{\phi}},\widetilde{\vect{\Gamma}}_{\vect{\phi}} )^{\otimes N}} \right|$. By the Cauchy--Schwarz inequality,
\begin{equation}
\left|\bra{(\widetilde{\vect{\nu}}_{\vect{\varphi}},\widetilde{\vect{\Gamma}}_{\vect{\varphi}} )^{\otimes N}}\Pi_{\vect{n}'}\ket{(\widetilde{\vect{\nu}}_{\vect{\phi}},\widetilde{\vect{\Gamma}}_{\vect{\phi}} )^{\otimes N}} \right| \leq   \left\| \Pi_{\vect{n}'}\ket{
 (\widetilde{\vect{\nu}}_{\vect{\varphi}}, \widetilde{\vect{\Gamma}}_{\vect{\varphi}})^{\otimes N}
 } \right\|  \left\| \Pi_{\vect{n}'}\ket{
 (\widetilde{\vect{\nu}}_{\vect{\phi}}, \widetilde{\vect{\Gamma}}_{\vect{\phi}})^{\otimes N}
 } \right\|. 
\end{equation}
Since $\bra{({\vect{\nu}}, {\vect{\Gamma}})^{\otimes N} }\Pi_{\vect{n}}\ket{
 ({\vect{\nu}}, {\vect{\Gamma}})^{\otimes N}
 } = \left\| \Pi_{\vect{n}}\ket{
 ({\vect{\nu}}, {\vect{\Gamma}})^{\otimes N}
 } \right\|^2 = \Mult(\vect{n}|\vect{\nu})
    $, we have the bound
\begin{equation}
\label{eq:magboundmulti}
\left|\bra{(\widetilde{\vect{\nu}}_{\vect{\varphi}},\widetilde{\vect{\Gamma}}_{\vect{\varphi}} )^{\otimes N}}\Pi_{\vect{n}'}\ket{(\widetilde{\vect{\nu}}_{\vect{\phi}},\widetilde{\vect{\Gamma}}_{\vect{\phi}} )^{\otimes N}} \right| \leq \sqrt{ \Mult(\vect{n}'| \widetilde{\vect \nu}_{\vect{\varphi}})\Mult(\vect{n}'| \widetilde{\vect \nu}_{\vect{\phi}})}.
    \end{equation}
Upper-bounding the multinomials using \eqref{eq:typebound} then gives an exponential bound in $N$, whose rate exponent is the average of the relative entropies between the normalized final occupations $\ov{n}' = \vect{n}'/N$ and the scattered virtual weights $\widetilde{\vect{\nu}}_{\vect{\phi}}$ and $\widetilde{\vect{\nu}}_{\vect{\varphi}}$:
\begin{equation}
\label{eq:integrandbound}
\left|\bra{(\widetilde{\vect{\nu}}_{\vect{\varphi}},\widetilde{\vect{\Gamma}}_{\vect{\varphi}} )^{\otimes N}}\Pi_{\vect{n}'}\ket{(\widetilde{\vect{\nu}}_{\vect{\phi}},\widetilde{\vect{\Gamma}}_{\vect{\phi}} )^{\otimes N}} \right| \leq e^{-\frac{N}{2}\left[ D(\ov{n}'\|\widetilde{\vect{\nu}}_{\vect{\varphi}}) + D(\ov{n}'\|\widetilde{\vect{\nu}}_{\vect{\phi}}) \right] }.
\end{equation}
Since the relative entropy is strictly positive unless its two arguments coincide, the integrand is  exponentially small at points  $(\vect{\phi}, \vect{\varphi})$ for which $\widetilde{\vect{\nu}}_{\vect{\phi}}  $ or $\widetilde{\vect{\nu}}_{\vect{\varphi}}$ differ from  $\ov{n}'$.

The bound \eqref{eq:integrandbound} has two important consequences for the asymptotic analysis. First, if for a given $\vect{n}'$ there is no $\vect{\phi} \in T$ with $\ov{n}'=\widetilde{\vect{\nu}}_{\vect{\phi}}$, the integral in \eqref{eq:probtorus} is exponentially small in magnitude, and hence so is $P(\vect{n}'|U,\vect{n}, \vect{\Gamma})$, since the choice $\vect{\nu}= \ov{n}$ guarantees that $\Mult(\vect{n}|\vect{\nu})$ scales only polynomially with $N$. Second, if $\vect{n}'$ is such that $P(\vect{n}'|U,\vect{n}, \vect{\Gamma})$ is not exponentially small, then only contributions to \eqref{eq:probtorus} from points $(\vect{\phi}, \vect{\varphi})$ with $\ov{n}'=\widetilde{\vect{\nu}}_{\vect{\phi}} = \widetilde{\vect{\nu}}_{\vect{\varphi}}$ need be considered, all others being exponentially small.

\section{Semiclassical asymptotics}\label{sec:asy}

\subsection{The classical map and the classical measure}

The asymptotic analysis of the MSP relies on a direct correspondence with a ``classical'' version of our $M$-port interferometer, in which a source of classical coherent light of total intensity $I$ is split into $M$ input beams of intensities $\vect{I} = (I_1, \ldots, I_M)$ such that the relative intensities $\vect{I}/I$ match the normalized initial occupations $\ov{n} = \vect{n}/N$ of $\ket{\vect{n}, \vect{\Gamma}}$. Before entering the interferometer, the internal degree of freedom in each beam is manipulated and the relative phases rotated to $\vect{\phi}$, so that the light entering the interferometer is a coherent state of $\langle N \rangle \propto I$ photons, all in the single-particle state $\ket{ \vect{\nu}_{\vect{\phi}}, \vect{\Gamma}_{\vect{\phi}}}$ of the toroidal expansion, Eq.~\eqref{eq:vsrot}.
Once scattered, the light exits with each photon transformed to the scattered virtual state $\ket{ \widetilde{\vect{\nu}}_{\vect{\phi}},\widetilde{\vect{\Gamma}}_{\vect{\phi}} }$ of Eq.~\eqref{eq:defscatvs}. Photodetectors at the output ports record the outgoing intensities $\vect{I}'$, which, normalized, define a point $\vect{\nu}' =\vect{I}'/I$ on the  simplex of relative output intensities. In the classical setting, $\langle N \rangle \gg1$ and the relative intensities become deterministic and from \eqref{eq:clasmapdef} are given by
\begin{equation}
\label{eq:map}
\vect{\nu}' =\widetilde{\vect\nu}_{\vect\phi}=\Diag\!\big(U\,t_{\vect\phi}\,\margvsn\,t_{\vect\phi}^\dagger U^\dagger\big), \qquad \margvsn := [\ov{n}]^{1/2}\, \vect{\Gamma}\,  [\ov{n}]^{1/2}.
\end{equation}
Equation~\eqref{eq:map} thus defines a smooth map $\widetilde{\vect{\nu}}:T\rightarrow\Delta$ from the torus of relative phases to the probability simplex of final normalized classical intensities. We refer to this map as the \emph{classical map}.

\begin{figure}[t]
 \includegraphics[width=\columnwidth]{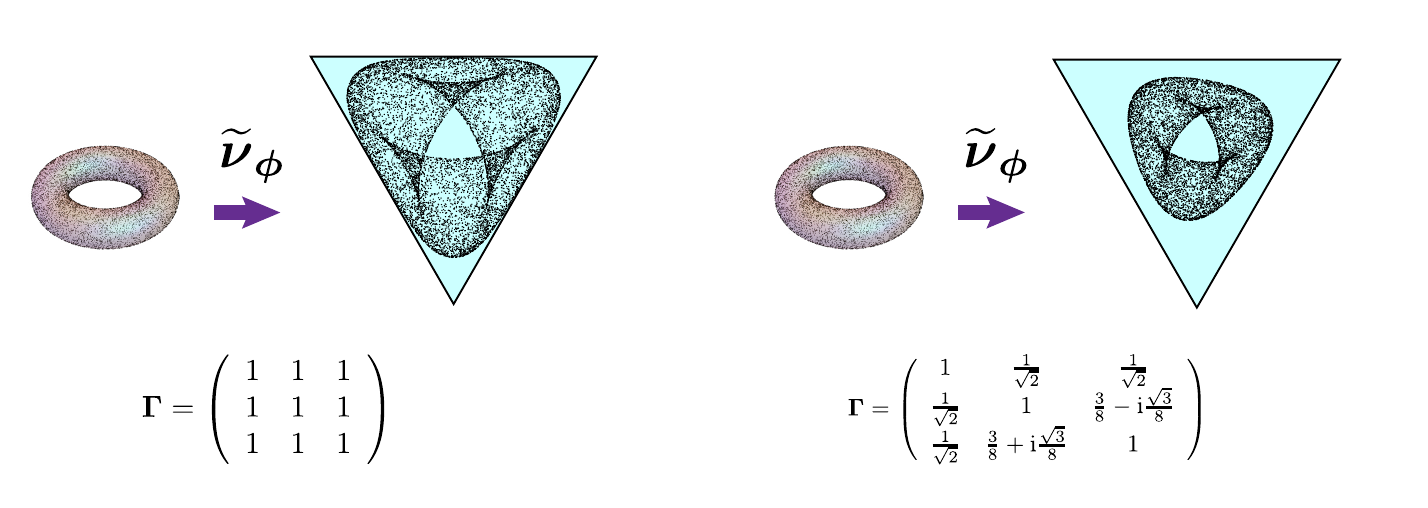}
\caption{The classical map \eqref{eq:map} for a three-port tritter with $\ov{n}=(0.7,0.25,0.05)$, for two degrees of indistinguishability: perfect indistinguishability (left) and partial indistinguishability (right), according to the indicated distinguishability matrices. The map is represented by the image on the simplex of a random collection of points on the torus. The image of the torus is the classically allowed region $\mathcal{R}$, which may entirely surround voids (points with no preimage on the torus). As the off-diagonal entries of $\vect{\Gamma}$ deviate from one, the size and shape of $\mathcal{R}$ can change qualitatively. The classical measure describes the density of points on the simplex 
and diverges at the creases of the map, where the Jacobian of $\tilde{\vect\nu}_{\vect{\phi}}$ vanishes.  }
\label{fig:map}
\end{figure}

The correspondence between the asymptotic analysis and the classical setup rests on two features of this classical map. The first is the \emph{classically allowed region} $\mathcal{R}=\widetilde{\vect\nu}(T)$, the image of the torus under the classical map. Geometrically, $\mathcal{R}$ can be pictured as an $(M-1)$-torus in the $(M^2-1)$-dimensional space of port density matrices, deformed by the action of $U$ and projected onto the diagonal elements of the density matrix---or, more graphically, ``squashed'' onto the probability simplex $\Delta$ (Fig.~\ref{fig:map}). Because the simplex is simply connected while the torus is not, the map is generically many-to-one: a typical point $\vect{\nu}'\in\mathcal{R}$ has a preimage $\vect\phi(\vect \nu')$ consisting of several discrete branches, and the number of branches changes across $\mathcal{R}$. Where two branches merge, the Jacobian determinant $\det (\partial\widetilde{\vect\nu}/\partial\vect\phi)$ vanishes; these ``creases''---the images of the points on the torus where the determinant vanishes--- are a multiport analog of the turning points of a Hamiltonian classical system.

The second feature underlying the asymptotic analysis is a probability distribution on the simplex of relative intensities that we call the \emph{classical measure}. Imagine repeating, many times, an experiment in which the initial relative phases $\vect{\phi}$ are selected randomly and uniformly on the torus, all other parameters fixed, and recording the resulting normalized output intensities $\vect{\nu}'$ on the probability simplex (Fig.~\ref{fig:map}). In the limit of infinite repetitions, the resulting (normalized) density of points on the simplex is the classical measure (or, more precisely, the classical pushforward measure)---the image of the uniform probability measure on the torus under the classical map. At a given $\overline{\vect \nu}'$, this measure sums contributions from all preimages of $\vect{\nu}'$, i.e., the angles $\vect{\phi}_\alpha$ such that $\widetilde{\vect{\nu}}_{\vect{\phi}_\alpha}=\vect{\nu}'$:
\begin{equation}
\label{eq:pushfwd}
f_{\mathrm{clas}}(\vect{\nu}'|U,\ov{n},\vect{\Gamma})=\frac{\mathbf{1}_{\mathcal{R}}(\vect \nu')}{(2\pi)^{M-1}}
\sum_{\vect\phi_\alpha\in\vect\phi(\vect \nu')}\left|\det\left(\frac{\partial\widetilde{\vect\nu}_{\vect{\phi}}}{\partial\vect\phi}\right)\right|^{-1}_{\vect\phi_\alpha},
\end{equation}
where $\mathbf{1}_{\mathcal{R}}(\vect \nu')$ is the indicator function of the classical region, and the Jacobian determinant is evaluated at the corresponding preimage angle $\vect{\phi}_\alpha$. The density thus diverges where this determinant vanishes, i.e., at the creases of $\mathcal{R}$.

As we show next, the classical region and the classical measure determine two key features of the asymptotic  MSP on the probability simplex, expressed in terms of normalized output occupations $\overline{\vect{n}}' =\vect{n}'/N$. The first is that the MSP  is effectively supported on the classical region $\mathcal{R}$, with exponentially small probability in the complementary, classically forbidden region $\overline{\mathcal{R}}$. The second is that the slowly-varying ``semiclassical'' envelope of the MSP is asymptotically given by the classical measure.

\subsection{Effective support of the MSP}\label{sec:effectivesupport}

Our first main asymptotic result concerns the effective support of the MSP $P(\vect{n}'|U,\vect{n}, \vect{\Gamma})$. If a final occupation $\vect{n}'$ is classically forbidden, i.e., $\overline{\vect{n}}' \notin \mathcal{R}$, then $P(\vect{n}'|U,\vect{n}, \vect{\Gamma})$ is exponentially small in $N$. More precisely:
\begin{theorem}
\label{thm:support}
For any $\vect{n}'$ in the occupation simplex, the multiphoton scattering probability $P(\vect{n}'|U,\vect{n}, \vect{\Gamma})$ can be bounded by
\begin{equation}
\label{eq:supportthm1}
    P(\vect{n}'|U,\vect{n}, \vect{\Gamma}) \leq   \,{(N+1)}^{M}\,e^{-N D(\ov{n}'\|\mathcal{R})},
\end{equation}
where 
\begin{equation}
    D(\ov{n}'\|\mathcal{R}) := \min_{\vect{\nu}' \in \mathcal{R}} D(\ov{n}'\|\vect{\nu}')
\end{equation} 
is the minimum relative entropy from $\ov{n}'$ to any point in $\mathcal{R}$.
\end{theorem}
This result follows from bounding the integral in the toroidal expansion of the MSP \eqref{eq:probtorus} by the maximal absolute value of the integrand, which from \eqref{eq:magboundmulti} is
\begin{equation}
    P(\vect{n}'|U,\vect{n}, \vect{\Gamma}) \leq  \max_{\vect{\phi}\in T } \frac{\Mult(\vect{n}'|\widetilde{\vect\nu}_{\vect \phi})}{\Mult(\vect{n}|\ov{n})}   .
\end{equation}
This is obtained by applying \eqref{eq:typebound} to upper-bound the numerator and lower-bound the denominator. Note that, as a consequence of the choice $\vect{\nu}_{\vect{\phi}}= \ov{n}$, the multinomial $\Mult(\vect{n}|\ov{n})$ contributes only a polynomial prefactor to the bound in \eqref{eq:supportthm1}.

As a consequence of the theorem, the occupation simplex splits into ``classically allowed'' ($\ov{n}' \in \mathcal{R}$) and ``classically forbidden'' ($\ov{n}' \notin \mathcal{R}$) regions, with exponentially small MSP in the latter. Interestingly, for $M \geq 3$ the classically forbidden region may include voids enclosed by $\mathcal{R}$, arising from the image of the ``donut hole'' of the torus; these are discussed further in Sec.~\ref{sec:bunching}. Since the number of points in the occupation simplex scales as $N^{M-1}$, the \emph{total} probability of all points $\vect{n}$ for which $\ov{n}'$ is a fixed distance $\epsilon$ from $\mathcal{R}$ also vanishes exponentially fast as $N \rightarrow \infty$ with $\ov{n}$ fixed. Thus, asymptotically, the MSP is effectively supported on the classical region.

\subsection{Decomposition of the MSP in the classical region} 

Our next result concerns the asymptotic behavior of the MSP in the interior of the classically allowed region. We restrict this analysis to \emph{regular points} of $\mathcal{R}$---those not lying on the creases of the classical map. Precisely, $\vect{\nu}'$ is a regular point in $\mathcal{R}$ if the preimage set $\vect{\phi}(\vect{\nu}') = \{ \vect{\phi} \in T \,|\, \widetilde{\vect{\nu}}_{\vect{\phi}} =\vect{\nu}'\}$ is discrete and $\det \frac{\partial \widetilde{\vect{\nu}}}{\partial{\vect{\phi}}} \neq  0$ in a sufficiently small neighborhood of each preimage $\vect{\phi}_\alpha \in \vect{\phi}(\vect{\nu}')$. A point $\vect{n}'$ in the occupation simplex is understood to be regular if $\ov{n}'$ is a regular point of the classical map.

For a regular point $\vect{n}'$, only contributions from regions of $T \times T$ where $\ov{n}' \simeq \widetilde{\vect{\nu}}_{\vect{\phi}} \simeq \widetilde{\vect{\nu}}_{\vect{\varphi}}$ are relevant to the integral \eqref{eq:probtorus} for $P(\vect{n}'|U,\vect{n}, \vect{\Gamma})$. To formalize this, note that since $\ov{n}'$ is regular, we can define, for a sufficiently small  $\epsilon >0$,  a ``relative-entropy ball'' $S(\ov{n}', \epsilon)$ of regular points around $\ov{n}'$ such that: (1) for any $\vect{\nu}' \in S(\ov{n}', \epsilon)$, $D(\ov{n}'\| \vect{\nu}') < 2\epsilon$; and (2) the preimage set of $S(\ov{n}', \epsilon)$ under the classical map breaks up into disjoint sets $S^{-1}_\alpha(\ov{n}', \epsilon) \subset T$, each containing exactly one preimage $\vect{\phi}_\alpha \in \vect{\phi}(\ov{n}')$ of $\ov{n}'$. Thus, the double integral in the toroidal expansion \eqref{eq:probtorus} for $P(\vect{n}'|U,\vect{n}, \vect{\Gamma})$ splits into a sum of integrals over the domains defined by the preimage sets $S^{-1}_\alpha$, plus an error term; schematically, denoting the full integral by $I$,
\begin{equation}
\label{eq:intdeco}
    I = \sum_{\alpha} I_{\alpha, \alpha} + \sum_{\alpha \neq \beta} I_{\alpha, \beta} +I_\text{err},
\end{equation}
where $I_{\alpha, \beta}$ is the same integral as in \eqref{eq:probtorus}, restricted to $S^{-1}_\alpha(\ov{n}', \epsilon) \times S^{-1}_\beta(\ov{n}', \epsilon)$, and $I_\text{err}$ is the integral over the remaining domains. From the definition of $S(\ov{n}', \epsilon)$ and the bound \eqref{eq:integrandbound}, the error term satisfies $|I_\text{err}| \leq e^{- N \epsilon}$. The MSP for a regular point $\vect{n}'$ can then be decomposed as
\begin{equation}
\label{eq:MSPdeco}
    P(\vect{n}'| U, \vect{n}, \vect{\Gamma}) = P_{\text{slow}}(\vect{n}'| U, \vect{n}, \vect{\Gamma}) + \delta P(\vect{n}'| U, \vect{n}, \vect{\Gamma}) + o((N+1)^M e^{-N \epsilon}),
\end{equation}
where $P_{\text{slow}}(\vect{n}'| U, \vect{n}, \vect{\Gamma})$ is a slowly-varying term from the ``diagonal'' integrals $I_{\alpha, \alpha}$ in \eqref{eq:intdeco} and $\delta P(\vect{n}'| U, \vect{n}, \vect{\Gamma})$ is the contribution from the ``off-diagonal'' integrals $I_{\alpha,\beta}$ with $\alpha \neq \beta$.  The $(N+1)^M$ prefactor in the  error term comes  from the polynomial behavior of $\Mult(\vect{n}|\ov{n} )$; with an arbitrarily small but finite value of $\epsilon$, the error term decays exponentially fast as $N \rightarrow \infty$.

Equation \eqref{eq:MSPdeco}  mirrors, for the toroidal expansion, the decomposition \eqref{eq:tpeinterf} of a discrete tensor power expansion. The first term arises from regions where $\vect{\phi} \simeq \vect{\varphi}$, and behaves approximately as a mixture of multinomials centered at $\vect{n}'= N \widetilde{\nu}_{\vect{\phi}}$, varying over the occupation simplex at the ``slow'' scale of order $N$. The second term involves integrals $I_{\alpha,\beta}$ from finitely separated regions, with $\vect{\phi}- \vect{\varphi} \simeq \vect{\phi}_\alpha - \vect{\phi}_\beta$, and, from the phases of \eqref{eq:probtorus}, produces oscillations at the ``fast'' $O(1)$ scale of the point spacing, averaging out (up to exponentially small errors) at the scale of the slow contribution. The term $\delta P(\vect{n}'| U, \vect{n}, \vect{\Gamma})$ therefore accounts for possible interference fringes in the MSP, and can be regarded as the ``interference term''.

\subsection{Asymptotics of the slowly-varying density} 

To discuss the asymptotic behavior of the slowly-varying term, it is convenient to work with normalized initial and final occupations $\ov{n}$ and $\ov{n}'$, and to rescale the MSP  so that as $N \rightarrow \infty$ the resulting function can be interpreted as a density on the probability simplex. We thus define
\begin{equation}
    f_\text{slow}(\ov{n}'|U, \ov{n}, \vect{\Gamma}) := N^{M-1} P_\text{slow}(\vect{n}'|U, \vect{n}, \vect{\Gamma}).
\end{equation}
The asymptotic limit of this density is then understood as $\lim_{N \rightarrow \infty}f_\text{slow}(\ov{n}'|U, \ov{n}, \vect{\Gamma})$, with $\ov{n}'$ and $\ov{n}$ held fixed.

To obtain an expression amenable to asymptotic analysis, we write the integrand of \eqref{eq:probtorus} as
\begin{equation}
e^{-\ii \vect{n}\cdot (\vect{\phi} - \vect{\varphi})} \bra{(\widetilde{\vect{\nu}}_{\vect{\varphi}},\widetilde{\vect{\Gamma}}_{\vect{\varphi}} )^{\otimes N}}\Pi_{\vect{n}'}\ket{(\widetilde{\vect{\nu}}_{\vect{\phi}},\widetilde{\vect{\Gamma}}_{\vect{\phi}} )^{\otimes N}} = \Mult(\vect{n}'|\ov{n}') e^{-N\Lambda(\vect\Phi) },
\end{equation}
where, for integrals on $T \times T$, we let $\vect{\Phi}=(\vect{\phi},\vect{\varphi})$ and define
\begin{equation}
\label{deflambda}
\Lambda(\vect{\Phi})  : =  \ii   
\ov{n}\cdot( \vect{\phi} - \vect{\varphi}) -     \sum_{k=1}^M \overline{n}_k'   \ln \frac{{\xi}_k(\vect{\Phi}) }{\overline{n}_k'},   \qquad {\xi}_k(\vect{\Phi}) := \bra{k} U t_{\vect{\phi}}\, \margvsn t_{\vect{\varphi}}^{\dagger} U^\dagger \ket{k}.
\end{equation}
Letting $d\mu(\vect{\Phi}) = d\mu(\vect{\phi})d\mu(\vect{\varphi})$, the slowly-varying density can be written as
\begin{equation}
\label{eq:slowasympt}
    f_\text{slow}(\ov{n}'|U, \ov{n}, \vect{\Gamma}) =   \frac{ \Mult(\vect{n}'|\ov{n}')}{\Mult(\vect{n}|\ov{n})}  \sum_{S^{-1}_\alpha} \overline{I}_{\alpha,\alpha}, \qquad  \overline{I}_{\alpha,\alpha} := N^{M-1}\int_{S_\alpha^{-1}\times S_\alpha^{-1}}d\mu(\vect\Phi)\,e^{-N\Lambda(\vect\Phi)},
\end{equation}
where, as before, $S^{-1}_\alpha$ denotes the preimage set $S^{-1}_\alpha(\ov{n}', \epsilon)$ of the relative-entropy ball around $\ov{n}'$ of radius $2\epsilon$, containing exactly one preimage $\vect{\phi}_\alpha \in \vect{\phi}(\ov{n}')$ of $\ov{n}'$. At each diagonal point $\vect{\Phi}_\alpha := (\vect{\phi}_\alpha,\vect{\phi}_\alpha)$, the exponent $\Lambda(\vect\Phi)$ vanishes, since $\vect{\xi}(\vect{\Phi}_\alpha) = \ov{n}'$ by construction, and is stationary there, as can be verified from the partial derivatives of $\Lambda$ with respect to $\phi_l$ and $\phi_l'$, which for general $\vect{\Phi}$ are
\begin{eqnarray}
\label{eq:firstderivs}
\frac{\partial \Lambda}{\partial \phi_l} & = &  \ii\, \left(\overline{n}_l -  \sum_{k=1}^M \frac{\overline{n}_k'}{\xi_k(\vect{\Phi})} \bra{l}\, \margvsn \, t_{\vect{\varphi}}^\dagger U^\dagger \ket{k}\bra{k}  U t_{\vect{\phi}}\ket{l} \right),  \\
 \frac{\partial \Lambda}{\partial \varphi_l} & = &  \frac{1}{\ii}\left( \overline{n}_l - \sum_{k=1}^M \frac{\overline{n}_k'}{\xi_k(\vect{\Phi})}\bra{l}  t_{\vect{\varphi}}^\dagger U^\dagger \ket{k}\bra{k}  U t_{\vect{\phi}}\, \margvsn\, \ket{l} \right).
\end{eqnarray}
At $\vect{\Phi}_\alpha$, the sums yield the diagonal elements $\bra{l} \margvsn\ket{l} = \overline{n}_l$, so the derivatives vanish.

The vanishing of $\Lambda$ and its derivatives at $\vect{\Phi}_\alpha$ means the integral $\overline{I}_{\alpha, \alpha}$ is dominated by the vicinity of $\vect{\Phi}_\alpha$, where the integrand is $O(1)$ and stationary, with contributions away from this region decaying exponentially fast, per the bound \eqref{eq:integrandbound}. As $N \rightarrow \infty$, the region of stationarity shrinks around $\vect{\Phi}_\alpha$ and the integral is increasingly well approximated by the (complex) Laplace method, i.e., the Gaussian integral from the second-order Taylor expansion of $\Lambda(\vect{\Phi})$ around $\vect{\Phi}_\alpha$. Thus, asymptotically,
\begin{equation}
\label{eq:limintalpha}
\lim_{N\rightarrow\infty}
\overline{I}_{\alpha,\alpha}
=\frac{1}{(2\pi)^{M-1}
\sqrt{\det H_{\alpha,\alpha}}
},
\end{equation}
where $H_{\alpha, \alpha}$ is the Hessian of $\Lambda$ at $\vect{\Phi}_\alpha$. For the factor $\Mult(\vect{n}'|\ov{n}')/\Mult(\vect{n}|\ov{n})$ in \eqref{eq:slowasympt}, the Stirling bound $
 e^\frac{1}{12 n+1} \leq \frac{ n! e^{n} }{ \sqrt{2 \pi} n^{n+ 1/2}} \leq e^\frac{1}{12 n}
$ gives the asymptotic limit
\begin{equation}
\label{eq:asymptmultifact}
\lim_{N \rightarrow \infty} \frac{ \mathcal{M}(\vect{n}'|\ov{n}')}{\mathcal{M}(\vect{n}|\ov{n})} = \sqrt{\prod_{i=1}^{M}\frac{\overline{n}_i}{\overline{n}_i'}} .
\end{equation}
The computation of the Hessian determinant in \eqref{eq:limintalpha} is straightforward but lengthy; the details can be found in Appendix~\ref{hessappendix}. The result is that the Hessian is proportional to the square of the Jacobian determinant of the classical map, with proportionality constant given by the square of the asymptotic limit \eqref{eq:asymptmultifact}:
\begin{equation}
    \label{hessrel}
     \det{H}_{\alpha, \alpha} = \left(\prod_{i=1}^{M}\frac{\overline{n}_i}{\overline{n}_i'}\right) \left|\det\left(\frac{\partial\widetilde{\vect\nu}}{\partial\vect\phi}\right)\right|^{2}_{\vect\phi_\alpha} .
\end{equation}
Putting everything together, we obtain the second main asymptotic result of the paper:

\begin{theorem}\label{thm:asymptoticclassical} 
The asymptotic limit of the slowly-varying density $f_\text{slow}(\ov{n}'|U, \ov{n}, \vect{\Gamma})$ at a regular point $\ov{n}'$ is the classical measure:
\begin{equation}
    \lim_{N \rightarrow \infty}f_\text{slow}(\ov{n}'|U, \ov{n}, \vect{\Gamma}) = f_\text{clas}(\ov{n}'|U, \ov{n}, \vect{\Gamma}). 
\end{equation}
\end{theorem}
Together with the limiting behavior of the MSP in the classically forbidden region, this shows that  away from the creases of the classical map, the asymptotic slowly-varying density on the full probability simplex is the classical measure, regardless of whether the photons are perfectly or partially indistinguishable.

\begin{figure}[t]
\centering
\includegraphics[width=1\textwidth]{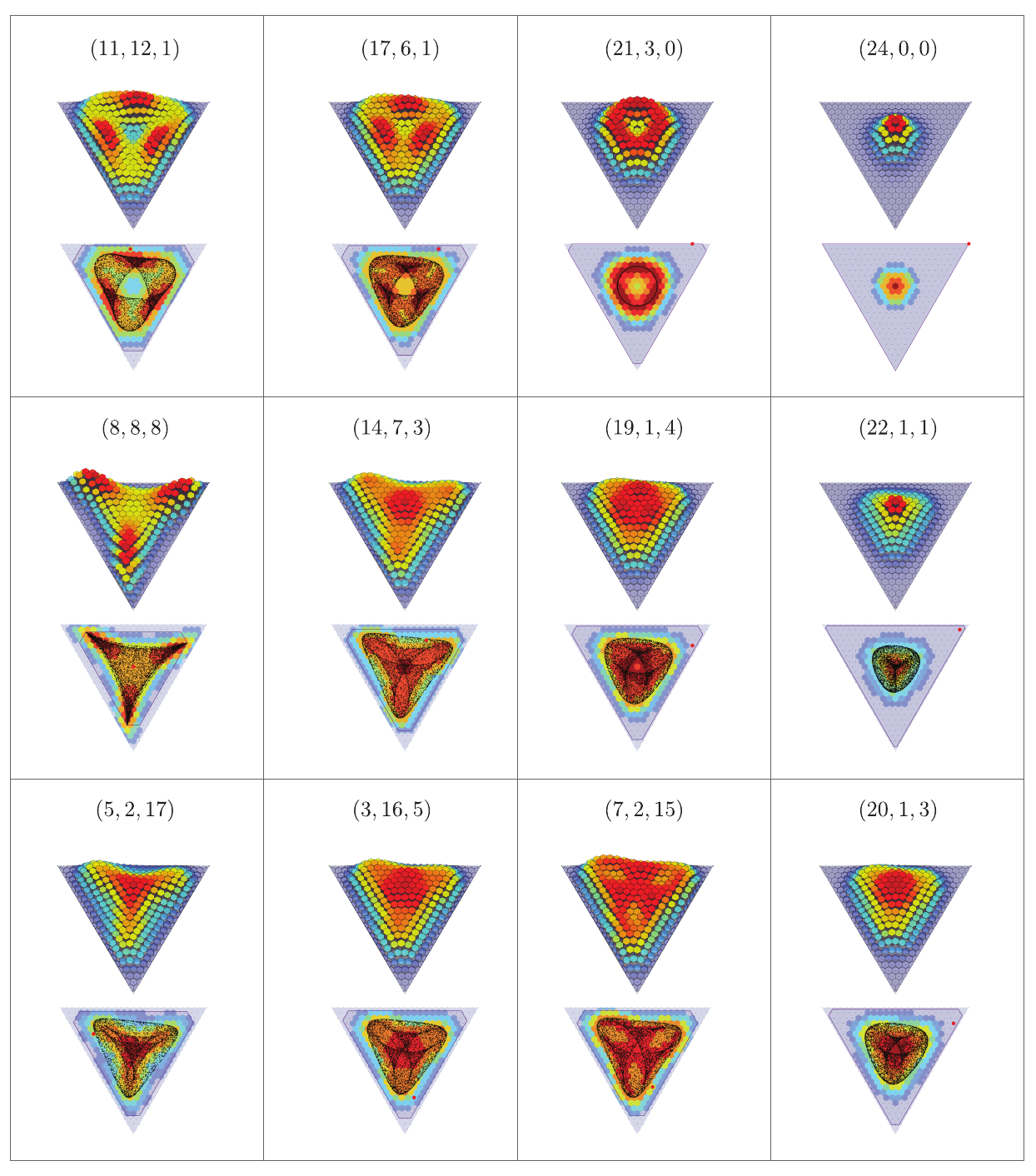}
\caption{The MSP  $P(\vect{n}'|U,\vect{n},\vect{\Gamma})$ on the normalized occupation simplex for $N=24$ partially distinguishable photons scattered by a tritter, for the same distinguishability matrix of \eqref{eq:map}
 and twelve different initial occupations $\vect{n}$ (box labels and red dots).  In each box,  the   MSP is  represented by heights and heat maps;  the bottom image compares the MSP with the image under the classical map of uniformly-sampled random torus points.   Also shown in the bottom images is the spectral permutahedron (see Section \ref{sec:bunching})--the convex hull of the permutations of the eigenvalues of $\mathlarger\margvsn$. 
 }
\label{fig:dist}
\end{figure}

\begin{figure}[t]
\centering
\includegraphics[width=\textwidth]{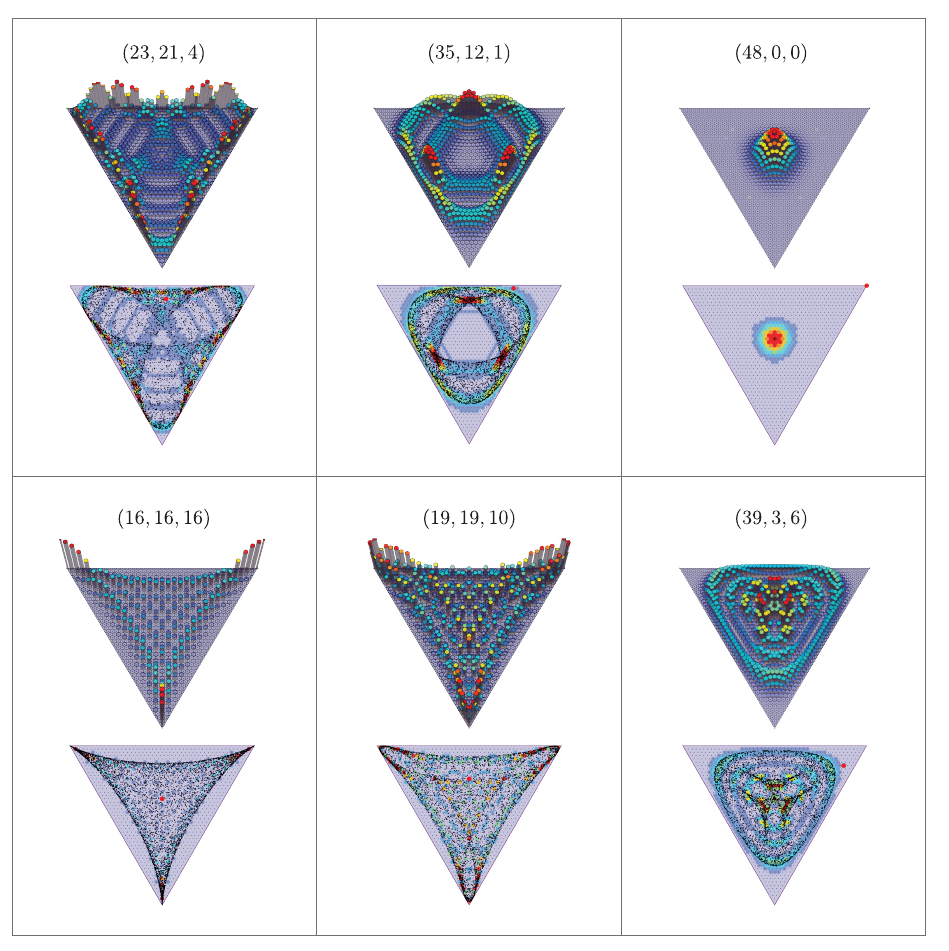}
\caption{As in Fig.~\ref{fig:dist} but for $N=48$ indistinguishable photons scattered by a tritter. In contrast to the previous case, the distributions show fast oscillations  modulated by the classical measure, in accordance with the semiclassical formula \eqref{eq:wkb}.  }
\label{fig:indist}
\end{figure}

The correspondence between the MSP and the classical measure is  evident even for relatively small values of $N$, as can be appreciated for the three-port tritter in Fig.~\ref{fig:dist} with $N=24$  
partially indistinguishable photons and 
and in Fig.~\ref{fig:indist} for $N=48$
perfectly indistinguishable photons, with the same indistinguishability matrices of Fig. \ref{fig:map}.  Both cases clearly show how the MSP accurately reflects the density of points of the classical map, including, in particular, regions of abnormally high probability near the creases of the map, where the classical measure diverges, as well as regions of very small probability at the voids of the classical map. The two figures also illustrate how  the ``fast'' oscillatory behavior of the MSP for perfectly indistinguishable photons is visibly suppressed for partially indistinguishable ones. As we now show, the asymptotic survival of interference fringes depends critically on the rank of $\vect{\Gamma}$.

\subsection{Large $N$ suppression of interference terms for full rank virtual states}

In addition to the slowly-varying term $P_\text{slow}(\vect{n}'|U, \vect{n}, \vect{\Gamma})$, Eq.~\eqref{eq:MSPdeco} allows for a rapidly oscillating interference term $\delta P(\vect{n}'|U, \vect{n}, \vect{\Gamma})$, which by \eqref{eq:intdeco} arises from the off-diagonal preimage-set pairs $S^{-1}_\alpha \times S^{-1}_\beta$ in $T \times T$. For such terms to persist as $N \rightarrow \infty$, however, the corresponding integrals $I_{\alpha, \beta}$ must be of the same order in $N$ as the diagonal integrals $I_{\alpha, \alpha}$. Using the definition \eqref{deflambda} of $\Lambda$, these off-diagonal integrals can be written as
\begin{equation}
    {I}_{\alpha,\beta} \propto  \int_{S_\alpha^{-1}\times S_\beta^{-1}}d\mu(\vect\Phi)\,e^{-N\Lambda(\vect\Phi)},
\end{equation}
with a proportionality factor that scales polynomially in $N$. A necessary, though not sufficient, condition for $I_{\alpha, \beta}$ to survive is thus that $\mathrm{Re}\,\Lambda=0$ be attained somewhere in the domain of integration, since otherwise the integral is exponentially suppressed.
Moreover, from the bound \eqref{eq:integrandbound}, this can only happen at $\vect \Phi_{\alpha,\beta}: = (\vect{\phi}_\alpha, \vect{\phi}_\beta)$. From the definitions of $\Lambda$ and the complex coefficients $\xi_k$, a sufficient condition for $\mathrm{Re}\,\Lambda=0$ is
\begin{equation}
\label{eq:offdiagcond} 
{\xi}_k(\vect{\Phi}_{\alpha, \beta}) = \bra{k} U t_{\vect{\phi}_\alpha}\, \margvsn \,    t_{\vect{\phi}_\beta}^{\dagger} U^\dagger \ket{k}  = \overline{n}'_k e^{\ii \theta_k},
\end{equation}
for some phase $\theta_k$. To see this is also necessary, note that the real part of $\Lambda$ can be written in terms of relative entropies as
\begin{equation}
\label{asymptoticrate}
\mathrm{Re}\Lambda =  D(\ov{n}'\| \vect{\zeta}) -\ln Z  ,
\end{equation}
where $\vect{\zeta}$ is the normalized probability distribution with $\zeta_k = |\xi_k|/Z$ and $Z=\sum_k|\xi_k|$ the normalization factor.
From the Cauchy--Schwarz inequality, $|{\xi}_k|   \leq \| \sqrt{\mathlarger\margvsn} t_{\vect{\varphi}}^{\dagger} U^\dagger \ket{k} \| \,  \, \| \sqrt{\mathlarger\margvsn} t_{\vect{\phi}}^{\dagger} U^\dagger \ket{k} \| = \sqrt{\widetilde{\vect \nu}_{\vect{\phi}}\widetilde{\vect \nu}_{\vect{\varphi}}}$, so $\ln Z \leq 0$ always. By the Gibbs inequality, $D(\ov{n}'\| \vect{\zeta}) >0$ unless $\vect{\zeta} = \ov{n}'$, in which case $Z=1$ and $|\xi_k| = \overline{n}_k'$.

It can then be shown that if $\mathlarger\margvsn$ is invertible, i.e., of rank $M$, \eqref{eq:offdiagcond} is impossible to satisfy except for trivial unitaries. To see this, define the unnormalized states
\begin{equation}
\ket{ \psi_k,\vect{\phi}}:= \sqrt{\margvsn}  \,  t^\dagger_{\vect{\phi}} U^\dagger \ket{k}.
\end{equation}
The condition $ \xi_k{(\vect{\Phi}_{\alpha, \beta})}=\overline{n}_k' e^{\ii \theta_k}$ then implies
\begin{equation}
\langle \psi_k, \vect{\phi}_\alpha| \psi_k, \vect{\phi}_\beta \rangle = \overline{n}_k' e^{\ii \theta_k}.
\end{equation}
Since $\vect{\phi}_\alpha, \vect{\phi}_\beta$ are preimages of $\ov{n}'$ under the classical map, it also holds that
\begin{equation}
    \langle \psi_k, \vect{\phi}_\beta| \psi_k, \vect{\phi}_\beta \rangle =  \langle \psi_k, \vect{\phi}_\alpha| \psi_k, \vect{\phi}_\alpha \rangle= \overline{n}_k'.
\end{equation}
Hence $| \psi_k, \vect{\phi}_\beta \rangle = e^{ \ii \theta_k}| \psi_k, \vect{\phi}_\alpha \rangle$ for every $k$ (including $\overline{n}'_k =0$, where both vectors vanish); therefore
$
\sqrt{\mathlarger\margvsn} t^\dagger_{\vect{\phi}_\beta} U^\dagger | k \rangle = \sqrt{\mathlarger\margvsn} t^\dagger_{\vect{\phi}_\alpha} U^\dagger e^{ \ii \theta_k}| k \rangle,
$
so
\begin{equation}
\label{thetatransf}
    \sqrt{\margvsn} = \sqrt{\margvsn}\, (  t_{\vect{\phi}_\alpha}^\dagger U^\dagger e^{\ii \Theta}  U t_{\vect{\phi}_\beta}  )    ,
\end{equation}
where $\Theta$ is the diagonal matrix with entries $\theta_k$. If $\mathlarger\margvsn$ is non-singular, dividing by $\sqrt{\mathlarger\margvsn}$ on both sides gives
\begin{equation}
U t_{\vect{\phi}_\alpha - \vect{\phi}_\beta} U^\dagger  =   e^{ \ii \Theta},
\end{equation}
so $e^{ \ii \Theta}$ and $t_{\vect{\phi}_\alpha - \vect{\phi}_\beta}$ must be isospectral diagonal matrices.
Hence the phases in $e^{ \ii \Theta}$ are a permutation of those in $t_{\vect{\phi}_\alpha - \vect{\phi}_\beta}$, and
$
 U t_{\vect{\phi}_\alpha - \vect{\phi}_\beta} U^\dagger  =   e^{ \ii \Theta}
$
can only be satisfied if $U$ is a permutation matrix.

It thus follows that if $\mathlarger\margvsn$ is of full rank, the only points on $T \times T$ where the coefficients $\xi_k$ attain the magnitudes $\overline{n}'_k$ needed for $\mathrm{Re}\,\Lambda (\vect{\Phi})$ to vanish are the diagonal points $(\vect{\phi}_\alpha, \vect{\phi}_\alpha)$. Our third main asymptotic result follows:
\begin{theorem}\label{thm:classicaldistribution}
For generic (non-permutation) $U$ and full-rank $\mathlarger\margvsn$, no off-diagonal interference terms survive in the asymptotic limit, in which case
\begin{equation}
    \lim_{N \rightarrow \infty} N^{M-1} P(N \ov{n}'| U, N \ov{n}, \vect{\Gamma}) = f_{clas}(\ov{n}'|U, \ov{n}, \vect{\Gamma}) ,
\end{equation}
for any $\ov{n}' \in \Delta$ that is either in the classically forbidden region or a regular point of $\mathcal{R}$. 
\end{theorem}
This result confirms the asymptotic behavior obtained in \cite{villalobosQuantumBeamSplitter2023} for the quantum beam splitter ($M=2$), and suggests that the absence of interference terms should be expected quite generally for partially indistinguishable photons, unless, perhaps, $\mathlarger\margvsn$ is rank-deficient. For non-zero normalized initial occupations $\overline{n}_k$, the rank of $\mathlarger\margvsn$ equals the rank of $\vect{\Gamma}$. Thus, a necessary condition for interference fringes to survive asymptotically is that $\vect{\Gamma}$ be rank-deficient. A general study of interference effects for such cases lies beyond the scope of this paper, but one prominent case deserves special consideration: perfectly indistinguishable photons, for which $\vect{\Gamma}$ is of rank one. As we show next, the toroidal expansion yields simple expressions for the phases underlying these interference patterns.

\subsection{The indistinguishable limit: a semi-classical formula for the transition amplitudes}
\label{sec:wkb}

In the case of perfectly indistinguishable photons, the MSP \eqref{eq:MSPindist} is the squared modulus of the transition amplitude $\bra{\vect{n}'}D^{(N)}(U)\ket{\vect{n}}$ in the $N$-particle Fock space $\mathcal{H}^{(N)}= \mathrm{Sym}_N( \mathcal{H}_p )$ associated to the port Hilbert space only. The toroidal tensor power expansion can then be adapted to obtain  an integral representation of $\bra{\vect{n}'}D^{(N)}(U)\ket{\vect{n}}$ on the torus of relative phases, well suited to asymptotic analysis. The resulting representation is equivalent to other integral representations already derived in the literature with similar aims \cite{shchesnovichASYMPTOTICEVALUATIONBOSONIC2013,englComplexScatteringCanonical2015}, and accordingly reproduces the asymptotic formulas obtained there through saddle-point methods. It is important to revisit these results from the toroidal perspective, since it furnishes a single-particle picture that illuminates the origin of the integral's stationary points and the resulting phases and prefactors of the amplitude. 

When $\Gamma_{ij}=1$, the single particle Hilbert space can be taken to be the port space $\mathcal{H}^{(1)}=\mathcal{H}_p \cong \mathbb{C}^M  $; the   virtual states in the toroidal expansion of  the initial Fock state $\ket{\vect{n}}$ and the scattered multiphotonic state  $D^{(N)}(U)\ket{\vect n}$ are then ordinary states in  $\mathbb{C}^M$. It then becomes convenient to adapt expressions \eqref{eq:vsrot} and \eqref{eq:scatvsrot} to virtual states parameterized  using virtual weights and phases, as  
\begin{equation}
 \ket{\vect{\nu},\vect{\phi}} = \sum_{k=1}^{M} \sqrt{\overline{n}_k} e^{\ii\phi_k}\ket{k},\qquad \ket{\widetilde{\vect{\nu}}_{\vect{\phi}},\widetilde{\vect{\phi}}_{\vect{\phi}}} = \sum_{k=1}^{M}\sqrt{\widetilde{\nu}_{\vect{\phi},k}}e^{\ii\widetilde{\phi}_{\vect{\phi},k}} \ket{k},
\end{equation}
where $\ket{\widetilde{\vect{\nu}}_{\vect{\phi}},\widetilde{\vect{\phi}}_{\vect{\phi}}} = U \ket{\vect{\nu},\vect{\phi}}$ so that
\begin{equation}
\label{eq:indistscatparam}
   \widetilde{{\nu}}_{\vect{\phi},k} = |\bra{k}U \ket{\vect{\nu},\vect{\phi}}|^2,  \qquad \widetilde{{\phi}}_{\vect{\phi},k} = \arg \bra{k}U \ket{\vect{\nu},\vect{\phi}}.
\end{equation}
As before, the initial virtual weights are set to the normalized occupations $\ov{n}$, and the initial phase vector $\vect{\phi}$ is understood with $\phi_1=0$.  For the scattered state, the phase  $\tilde{\phi}_1$  need not be zero, and the scattered virtual weights $\widetilde{{\vect\nu}}_{\vect{\phi}}$ are consistent with \eqref{eq:map} after the replacement 
$\mathlarger \margvsn \to \ket{\vect{\nu},\vect{\phi}} \bra{\vect{\nu},\vect{\phi}}{} \Big|_{\vect{\phi}=0} $.

The asymptotic formula for the transition amplitude $\bra{\vect{n}'}D^{(N)}(U)\ket{\vect{n}}$ is based on the two $U$-dependent functions of the initial angle $\vect{\phi}$ and the initial virtual weight $\vect{\nu}$ defined in \eqref{eq:indistscatparam}. To obtain an integral representation of the transition amplitude, we note that the complex coefficients $\xi_k(\vect{\Phi})$ in \eqref{deflambda} can be factored as
\begin{equation}
\label{eq:defz}
\xi_k(\vect{\Phi}) = \amp{k}{\widetilde{\vect{\nu}}_{\vect{\phi}},\widetilde{\vect{\phi}}_{\vect{\phi}}} \amp{\widetilde{\vect{\nu}}_{\vect{\varphi}},\widetilde{\vect{\phi}}_{\vect{\varphi}} }{k}.
\end{equation}
This factorization causes the double integral \eqref{eq:probtorus} to factor as the modulus squared of a single integral on $T$ which, up to combinatorial factors and an overall phase, equals $\bra{\vect{n}'}D^{(N)}(U)\ket{\vect{n}}$. The resulting integral representation takes the form
  \eqref{eq:slowasympt}:
\begin{equation}
\label{eq:torusindistamp}
\bra{\vect{n}'}D^{(N)}(U)\ket{\vect{n}} = \sqrt{\frac{ \Mult(\vect{n}'|\ov{n}')}{\Mult(\vect{n}|\ov{n})} }\int_T d\mu(\vect{\phi}) e^{ \ii N \mathlarger{\sigma}(\vect{\phi})  },
\end{equation}
where the exponent is a complex function that depends on the initial phases $\vect{\phi}$ and the $\vect{\phi}$-dependent parameters $\widetilde{\vect{\nu}}_{\vect{\phi}},\widetilde{\vect{\phi}}_{\vect{\phi}}$ of the scattered state; broken into real and imaginary parts, it is given by 
\begin{equation}
\label{eq:defsigma}
\sigma(\vect{\phi})
=
\ov{n}'\cdot\widetilde{\vect{\phi}}_{\vect{\phi}}
-\ov{n}\cdot\vect{\phi}
+\frac{\ii}{2}
D\!\left(
\ov{n}'\middle\|
\widetilde{\vect{\nu}}_{\vect{\phi}}
\right).
\end{equation}
Since the imaginary part is proportional to the relative entropy between $\ov{n}'$ and $\widetilde{\nu}_{\vect{\phi}}$, it vanishes for the values of $\vect{\phi}$ where  $\widetilde{\nu}_{\vect{\phi}} = \ov{n}'$, in other words, at the pre-images of $\ov{n}'$ under the classical map.  

As before, for large $N$ and regular $\ov{n}'$ in the classical region, the main contribution comes from the preimage points $\vect{\phi}_\alpha\in\vect{\phi}(\ov{n}')$ where $\widetilde{\vect{\nu}}_{\vect{\phi}_\alpha} = \ov{n}'$ and the integrand of \eqref{eq:torusindistamp} is magnitude one. That the integrand is also  stationary may be expected from \eqref{eq:firstderivs}, and is also shown in \ref{WKBappendix} in terms of the exponent $\sigma(\vect{\phi})$. Then, the complex Laplace method can be applied at each preimage $\vect{\phi}_\alpha$, yielding  the asymptotic expansion of the integral \eqref{eq:torusindistamp}:
\begin{equation}
\label{eq:laplaceindist}
\int_T d\mu(\vect{\phi}) e^{ \ii N \mathlarger{\sigma}(\vect{\phi})  } =  \frac{1}{(2\pi N)^{(M-1)/2}} \left[ \sum_{\vect{\phi}_\alpha \in \vect{\phi}(\ov{n}')}  \frac{e^{ \ii N \mathlarger{\sigma}(\vect{\phi}_\alpha)  }}{\sqrt{ \det(G_\alpha)}}   + O(1/N) \right],
\end{equation}
where $G_\alpha$ is the Hessian of $- \ii\sigma(\vect{\phi})$ at $\vect{\phi}_\alpha$, and, from \eqref{eq:defsigma}, $\sigma(\vect{\phi}_\alpha)$ are real phases.

Since the slowly-varying part of the squared modulus of this asymptotic amplitude must yield the classical measure, the slowly-varying prefactors  in \eqref{eq:laplaceindist} must have the form
\begin{equation}
\frac{1}{\sqrt{\det(G_\alpha)}}
=
\left(
\frac{\prod_k\overline n'_k}
{\prod_i\overline n_i}
\right)^{1/4}
\left|\det\left(\frac{\partial \widetilde{\vect{\nu}}_{\vect{\phi}} }{\partial\vect\phi}\right)\right|^{-1/2}_{\vect\phi_\alpha}
e^{\ii\delta_\alpha},
\end{equation}
where the $\delta_\alpha$ are $O(1)$ phases; this is shown in ~\ref{WKBappendix}, together with the  explicit values $\delta_\alpha$. The result is a WKB-like asymptotic formula for the transition amplitude of perfectly indistinguishable photons:
\begin{equation}
\label{eq:wkb}
\bra{\vect{n}'}D^{(N)}(U)\ket{\vect{n}} =\frac{1}{(2\pi N)^{M-1 \over 2}}
\left[ \sum_{\vect{\phi}_\alpha\in\vect{\phi}(\ov{n}')}
\left|\det\left(\frac{\partial \widetilde{\vect{\nu}}_{\vect{\phi}} }{\partial\vect\phi}\right)\right|^{-1/2}_{\vect\phi_\alpha}\!\!e^{\,\ii \left(N \sigma(\vect{\phi}_\alpha) + \delta_\alpha \right)}+O(1/N)\right],
\end{equation}
where the phases $\delta_\alpha$ are given by
\begin{equation}
\label{eq:action}
\delta_\alpha=
\left. \frac{1}{2} \sum_{k=1}^{M}
(\widetilde{\phi}_k-\phi_k)\right|_{\vect{\phi}_\alpha}
-
\frac{1}{2}\left. \arg\!\left[
\det\left(
\frac{\partial\widetilde{\vect{\nu}}_{\vect{\phi}}}
{\partial\vect{\phi}}
\right)
\right] \right|_{\vect{\phi}_\alpha}
-\frac{1}{2}\arg\det U-\frac{(M-1)\pi}{4}. 
\end{equation}
Thus, for indistinguishable photons, the virtual-state construction provides not only the slowly-varying envelope of the MSP through the classical map but also the phases governing the rapidly-oscillating interference pattern,
\begin{equation}
N \sigma(\vect{\phi}_\alpha)
=
\vect{n}'\cdot\widetilde{\vect{\phi}}_{\vect{\phi}_\alpha}
-
\vect{n}\cdot\vect{\phi}_\alpha,
\end{equation}
where $\widetilde{\vect{\phi}}_{\vect{\phi}_\alpha}$ are the phases of the scattered virtual state whose virtual weights match the final normalized occupations $\ov{n}'$.

As stated earlier, the formula in Eq.~\eqref{eq:wkb} is 
equivalent to known semiclassical formulas for bosonic 
transition amplitudes with perfectly indistinguishable 
photons, where the integral representations arise either 
from the large-$N$ continuum limit of a combinatorial 
formula for permanents in terms of contingency tables 
\cite{shchesnovichASYMPTOTICEVALUATIONBOSONIC2013}, or 
from coefficient extraction in the series expansion of an 
appropriate generating function 
\cite{englComplexScatteringCanonical2015}. In either 
case, the resulting integrals are evaluated by the 
saddle-point method, and for points in the classical region the 
saddle-point conditions reduce to single-particle 
conditions: finding the initial virtual state 
$\ket{\vect{\nu} =\ov{n}, \vect{\phi}}$ that, under $U$, 
maps to a scattered virtual state 
$\ket{\widetilde{\vect{\nu}}_{\vect{\phi}} =\ov{n}', 
\widetilde{\vect{\phi}}_{\vect{\phi}}}$.
 In particular, the equivalence with the formula in Eq.~(61) of Ref.~\cite{englComplexScatteringCanonical2015} is as 
 follows. Regarding $\vect{\nu}$ as canonically conjugate 
 momenta to the angles $\vect{\phi}$, the parameter 
 transformation $(\vect{\nu},\vect{\phi})\rightarrow (\widetilde{\vect{\nu}}, \widetilde{\vect{\phi}})$ 
 induced by $U$ is a canonical transformation; for 
 initial angles in branch $\alpha$, a type-4 generating 
 function for this transformation is $F_\alpha(\ov{n}, \ov{n}')= \sigma(\vect{\phi}_\alpha)$, when $\sigma(\vect{\phi}_\alpha)$ is expressed as a function of $\ov{n}$ and $\ov{n}'$. Since the initial angle $\vect{\phi}$ is the negative of the derivative of $F_\alpha$ with respect to $\ov{n}$, the Jacobian determinant of the classical map is related to the Van Vleck determinant of $F_\alpha$:
\begin{equation}
    \left|\det\left(\frac{\partial \widetilde{\vect{\nu}}_{\vect{\phi}} }{\partial\vect\phi}\right)\right|^{-1} = \left|\det\left(\frac{\partial^2 F_\alpha }{\partial\ov{n} \partial \ov{n}'}\right)\right|. 
\end{equation}
Thus, as in Ref.~\cite{englComplexScatteringCanonical2015}, the phase $N \sigma(\vect{\phi}_\alpha)$ and the Jacobian factor in \eqref{eq:wkb} can be expressed in terms of $F_\alpha$ and its Van Vleck determinant, respectively.  The Maslov phases of Ref.~\cite{englComplexScatteringCanonical2015} can likewise be understood as arising from a sign change in the Jacobian determinant when branches of the classical map meet at a crease. The toroidal formulation thus offers an alternative, closely related single-particle interpretation of the  known semiclassical formulas for perfectly indistinguishable photons.

\section{Photon bunching patterns and the classical map}
\label{sec:bunching}

The three main theorems of Sec.~\ref{sec:asy} translate directly into 
predictions about observable patterns of photon \emph{bunching} in 
multiphotonic scattering, i.e., the extent to which the MSP favors some 
final occupations $\vect{n}'$ over others. These patterns can be grouped 
into two categories of distinct origin. The first concerns patterns arising from the 
slowly-varying part of the MSP as a consequence of the geometry and topology 
of the classical map: these are \emph{voids} and \emph{caustics}, which 
follow from Theorems~\ref{thm:support} and~\ref{thm:asymptoticclassical} 
and apply regardless of the degree of indistinguishability. The second category
concerns rapidly oscillating interference fringes, which can drive the 
scattering probability to zero or near-zero at specific 
occupations---the interference nodes. This latter mechanism is the multiport 
generalization of the HOM-type total or near-total suppression arising 
from the destructive interference of multiphoton amplitudes
\cite{dittelTotallyDestructiveManyParticle2018,dittelTotallyDestructiveInterferencesymmetric2018,tichyZeroTransmissionLawMultiport2010,limGeneralizedHongOu2005}; 
by Theorem~\ref{thm:classicaldistribution}, such patterns survive asymptotically 
only when $\vect{\Gamma}$ is rank-deficient.

Voids correspond to final occupations $\vect{n}'$ with exponentially 
small probability when $\ov{n}'$ lies outside the classical region 
$\mathcal{R}$. Arising from a continuous map from the torus to the 
simplex, they are topological in origin. For $M=2$, the classical region 
is an interval and voids exist only outside its endpoints. For $M \geq 3$, 
a qualitatively new possibility appears: besides the region outside its 
outer edges, the classical region may enclose holes---the images of the 
``donut hole'' of the torus (Figs.~\ref{fig:map}, \ref{fig:dist}, 
and~\ref{fig:indist})---giving rise to voids that are enclaves within 
$\mathcal{R}$. For higher $M$, the morphology of these interior voids is 
expected to be richer still. The shape and location of the voids are 
sensitive to all conditioning parameters $\vect{n}$, $U$, and 
$\vect{\Gamma}$ of the MSP.

The second  type of  bunching pattern visible in the slowly-varying part of 
the MSP arises from the creases of $\mathcal{R}$, which we call 
\emph{caustics}, in analogy with ray-optics wavefront singularities: both 
arise from the same fold mechanism, whereby nearby points of a compact 
surface embedded in a higher-dimensional space map to the same, or nearly 
the same, point in a configuration space 
\cite{berryIVCatastropheOptics1980}. Caustics govern the complementary 
prediction to that of voids, namely \emph{enhanced} probability for 
certain occupations. Where 
$\det(\partial\widetilde{\vect\nu}/\partial\vect\phi)$ vanishes, branches 
of the preimage $\vect\phi(\ov{n}')$ merge and $f_\text{clas}$ diverges 
by Eq.~\eqref{eq:pushfwd}; the prediction is a ridge of anomalously high 
count probability tracing the crease locus, visible in 
Figs.~\ref{fig:dist} and~\ref{fig:indist} as sharp boundaries. For $M=2$, caustics reduce to 
turning points at the boundary of $\mathcal{R}$, delimiting the 
classically allowed and forbidden regions. For $M \geq 3$, a new phenomenon arises: caustics may lie in the interior of $\mathcal{R}$, delimiting subregions with 
different numbers of preimages of the classical map.

In addition to the bunching patterns set by the slowly-varying part of the MSP, 
rapidly-oscillating interference patterns are also possible when 
$\mathlarger\margvsn$ is rank-deficient, and can drive the scattering 
probability to zero or near-zero at specific occupations. For perfectly 
indistinguishable photons (Sec.~\ref{sec:wkb}), the interference fringes 
arise from the phases of the single-particle virtual states in the 
toroidal expansions of the initial and scattered multiphotonic states; in 
the semiclassical formula~\eqref{eq:wkb}, these are weighted by factors equal to 
the initial and final occupations. The toroidal expansion thus offers a 
convenient single-particle perspective on the bunching patterns associated 
with the MSP's interference fringes. Examples are shown in 
Fig.~\ref{fig:indist} for the balanced tritter, where the case of 
balanced initial occupations $\vect{n} = (N/3, N/3, N/3)$ shows what 
appears to be a three-port generalization of the even-odd structure known 
for the $M=2$ case with balanced inputs on a 50:50 beam splitter 
\cite{camposQuantummechanicalLosslessBeam1989}.

Finally, it is worth noting how indistinguishability shapes the size of the 
classical region. As can be appreciated in Fig.~\ref{fig:map}, the 
classical region tends to contract as $\vect{\Gamma}$ deviates from a 
matrix of all ones. This contraction can be linked to a ``spectral'' 
mechanism that constrains the size of $\mathcal{R}$, independently of the 
unitary $U$. Since $U t_{\vect\phi}$ is unitary for every $\vect\phi$, 
$\widetilde{\vect\nu}_{\vect\phi}$ is the diagonal of a matrix unitarily 
equivalent to $\mathlarger\margvsn$. By the Schur--Horn theorem 
\cite{schur1923UbereineKlasse1923,hornDoublyStochasticMatrices1954}, any 
such diagonal lies in the convex hull of the $M!$ permutations of the 
eigenvalues $\vect{\lambda} = (\lambda_1, \ldots, \lambda_M)$ of 
$\mathlarger\margvsn$---the \emph{spectral permutohedron} 
$\Pi(\vect\lambda)$ \cite{postnikovPermutohedraAssociahedra2009}---giving 
the containment
\begin{equation}
\label{eq:permhull}
\mathcal{R}=\widetilde{\vect\nu}(T)\;\subseteq\;\Pi\big(\vect\lambda\big),
\end{equation}
so that, for given $\vect{n}$ and $\vect{\Gamma}$, any region outside 
$\Pi(\vect\lambda)$ is a void for \emph{every} choice of $U$. For 
perfectly indistinguishable photons, the spectral permutohedron is the full 
simplex, but for partially indistinguishable photons it is a polyhedron 
that shrinks as the eigenvalues of $\mathlarger\margvsn$ become more 
similar, constraining the classical region accordingly. Examples are shown 
in Fig.~\ref{fig:dist}.

\section{Discussion and conclusion}

We have introduced a framework for the asymptotic analysis of multiphotonic 
scattering probabilities of partially indistinguishable photons in a 
lossless multiport interferometer. The framework applies to correlated mode 
states, a broad family of input states parametrized by the input occupations 
and the Gram matrix $\vect{\Gamma}$ of internal overlaps. These states are 
expanded in tensor powers of single-particle virtual states parametrized 
over a torus of relative phases; such tensor powers are $SU(M^2)$ coherent 
states and thus become classical in the limit $N\to\infty$. The scattered 
multiphotonic state can therefore be interpreted as a coherent superposition 
of $N$-photon coherent-state scattering experiments, each parametrized by a 
point on the torus. This interpretation leads to an integral representation 
of the scattering probability (Propositions 1 and 2), from which a 
classical map, from the torus to the probability simplex of output 
intensities, arises. The classical map governs the asymptotic behavior of the 
scattering probabilities in the large-photon limit.

Three main asymptotic results follow from this construction. When represented 
on the occupation simplex, the scattering probability concentrates on the 
classically allowed region---the image of the torus under the classical 
map---and is exponentially small outside it (Theorem 3). The slowly-varying 
part of the distribution converges to the classical measure induced by the 
map (Theorem 4). The rapidly oscillating interference terms are suppressed 
whenever $\vect{\Gamma}$ is of full rank (Theorem 5). The rank of 
$\vect{\Gamma}$ thus determines whether the asymptotic distribution is 
governed by the classical measure alone, or exhibits  multiphoton 
interference fringes. In the limiting case of perfectly indistinguishable 
photons, the framework also provides a single-particle interpretation of 
the known semiclassical formulas for the transition amplitudes.

For $M \geq 3$ ports, the classical map reveals two types of structure in 
the multiphotonic distribution that are distinct from the rapidly oscillating 
fringes characteristic of multiphoton interference. These structures arise 
from the topology of the classical map as a map from the torus to the 
simplex, have no counterpart in the two-port case, and are present 
regardless of the degree of indistinguishability. The first are 
voids---regions of exponentially suppressed probability enclosed within the 
classically allowed region. The second are caustics---ridges of enhanced 
probability due to creases of the map, which delimit subregions of the 
classically allowed region with different numbers of preimages. These 
features are already clearly identifiable at moderate photon numbers, as 
illustrated for $N=24$ partially distinguishable and $N=48$ 
indistinguishable photons in a $3\times 3$ optical network 
(Figs.~\ref{fig:dist} and \ref{fig:indist}).

Several extensions are natural within this framework. First, the bounds of 
Theorem 3 on the scattering probability in the classically forbidden region 
should be sharpened to yield the correct exponential rate functions; more 
broadly, a precise analysis of the transition from the classically forbidden 
to the classically allowed region, and of the behavior along the caustics 
where different subregions meet, would complete the asymptotic picture 
beyond the regular-point regime treated here. Second, the intermediate 
cases in which $\vect{\Gamma}$ is rank-deficient but not rank one, where 
interference terms may survive asymptotically, remain to be explored 
through explicit examples. Third, the connection with representation theory 
deserves further investigation. In the two-port case, the scattering 
probability admits a decomposition into irreducible representations 
dominated by a single dominant irrep 
\cite{villalobosQuantumBeamSplitter2023}; one may expect a similar structure 
in the general multiport case, suggesting that the classical map may be 
related to the group functions of the irreducible representations close to 
the dominant representation. Finally, we have illustrated our results 
throughout for $M=3$ ports; it would be of interest to explore the 
structural consequences of the classical map for higher $M$, where the 
topology of the classically allowed region may give rise to 
higher-dimensional analogs of voids and caustics.

\section*{Acknowledgments}

The work of HdG was supported by NSERC of Canada, by Lakehead University and by a grant from the Department
of Foreign Affairs, Trade and Development of the Government of Canada. AB acknowledges support from the Faculty of Science, Universidad de los Andes, project INV-2023-162-2814.

\appendix

\section{}
\label{TPprojappendix}
To prove \eqref{eq:proj}, let $a_{\psi}^\dagger$ be   the creation operator for the single particle mode $\ket{\psi} = \ket{\vect{\nu}, \vect{\gamma}}$
\begin{equation}
    a_{\psi}^\dagger = \sum_{i} \sqrt{\nu_i}a_{i, \chi_i}^\dagger, 
\end{equation}
in terms of the creation operators $a_{i, \chi_i}^\dagger$ defined in \eqref{eq:aichi}.  Then the state $e^{\lambda a_{\psi}^\dagger }\ket{\vect{0}}$, for arbitrary $\lambda$, can be expanded in either of two ways. The first one is 
\begin{align}
   e^{\lambda a_{\psi}^\dagger }\ket{0} & = \sum_{N'=0}^{\infty} \frac{(\lambda a_{\psi}^\dagger)^{N'}}{N'!}  \ket{\vect{0}} \\ 
   & = \sum_{N'=0}^{\infty} \frac{\lambda^{N'}}{\sqrt{N'!}}\ket{(\vect{\nu},\vect{\Gamma})^{\otimes {N'}}}, 
\end{align}
where in the last line we have used \eqref{eq:tps}. Alternatively, we can expand the exponent $e^{\lambda a_{\psi}^\dagger }\ket{0}$ in terms of the creation operators $a_{i, \chi}^\dagger$, and then expand the exponentials: 
\begin{align}
   e^{\lambda a_{\psi}^\dagger }\ket{\vect{0}} & = \sum_{n_1'=0}^{\infty} \ldots \sum_{n_M'=0}^{\infty} 
   \frac{(\lambda \sqrt{\nu_1}a_{(1,\chi_1)}^\dagger)^{n_1'}}{n_1'!} \ldots   \frac{(\lambda \sqrt{\nu_M} a_{(M,\chi_M)}^\dagger)^{n_M'}}{n_M'!} \ket{0} \\ 
   & = \sum_{n_1'=0}^{\infty} \ldots \sum_{n_M'=0}^{\infty}  
   \frac{\lambda^{n_1' + \ldots +n_M'}}{\sqrt{n_1'! \ldots n_M'!}} \sqrt{\nu_1}^{n_1'} \ldots  \sqrt{\nu_M}^{n_M'}\ket{\vect{n}',\vect{\Gamma}},
\end{align}
where in the last line we have used the definition \eqref{eq:cms} of the correlated mode states. Applying  the projection operator $\Pi_{\vect{n}}$ to the second expansion extracts the term  $\vect{n}'= \vect{n}$ in the sum, in which case the exponent $n_1' + \ldots + m_M'$ of $\lambda$ evaluates to $N$. Thus we have:
\begin{equation}  
   \frac{\lambda^{N}}{\sqrt{n_1! \ldots n_M!}}  \sqrt{\nu_1}^{n_1'} \ldots  \sqrt{\nu_M}^{n_M'} \ket{\vect{n}',\vect{\Gamma}} = \sum_{N'=0}^{\infty} \frac{\lambda^{N'}}{\sqrt{N'!}}\Pi_{\vect{n}}\ket{(\vect{\nu},\vect{\Gamma})^{\otimes {N'}}}.
\end{equation}
Since the expansion is true for all values of $\lambda$, it follows by comparing powers of $\lambda$ that only the term $N'= N$ survives on the right-hand side, in which case \eqref{eq:proj} follows.

\section{}
\label{hessappendix}
In this appendix we prove formula \eqref{hessrel} for the  determinant of the Hessian $H_{\alpha, \alpha}$ in the quadratic expansion
\begin{equation}
\label{quadlambda2}
\Lambda(\vect{\Phi}) =    \frac{1}{2}\delta\vect{\Phi}^T  H_{\alpha, \alpha} \delta\vect{\Phi}  + O(\delta\vect{\Phi}^3) 
\end{equation} 
around the diagonal point
 $\vect{\Phi}_{\alpha, \alpha} =  (\vect{\phi}_\alpha,\vect{\phi}_\alpha)$ for $ \vect{\phi}_\alpha \in  \vect{\phi}(\ov{n}')$, where $\xi_k(\vect{\Phi}_{\alpha, \alpha}) = \overline{n}_k'$ and $\Lambda(\vect{\Phi}_{\alpha, \alpha}) = 0$. Given that  the Gaussian integral that follows from this expansion is
 \begin{equation}
 \label{gaussintapp}
    \int d \mu(\vect{\Phi}) e^{- \frac{1}{2}\delta\vect{\Phi}^T  H_{\alpha \alpha} \delta\vect{\Phi} } = \left[(2 \pi)^{M-1} {\sqrt{ \det H_{\alpha, \alpha}}}\right]^{-1}, \, \qquad \vect{\Phi}=(\vect{\phi}, \vect{\varphi}), 
 \end{equation}
we shall obtain $\det H_{\alpha, \alpha}$  from the  integral directly after performing an appropriate change of variables. To this end,  we start with a volume-preserving transformation to new variables  $(\vect{\eta}, \vect{\varphi})$, where
\begin{equation}
\vect{\eta} = \vect{\phi} - \vect{\varphi}, \qquad \widehat{\vect{\phi}} = \frac{1}{2}(\vect{\phi}+ \vect{\varphi}).  
\end{equation}
The diagonal point  $\vect{\Phi}_{\alpha,\alpha} = (\vect{\phi}_\alpha,\vect{\phi}_\alpha)$ then corresponds to the values $\vect{\eta}=0, \widehat{\vect{\phi}}=\vect{\phi}_\alpha$.  Then,
\begin{equation}
    \xi_k(\vect{\eta}, \widehat{\vect{\phi}}) = \langle k| U e^{\frac{\ii}{2}\vect{h}\cdot \vect{\eta}}  \rho_{\widehat{\vect{\phi}}} e^{\frac{\ii}{2}\vect{h}\cdot \vect{\eta}} U^\dagger| k \rangle, 
\end{equation}
where $\vect{h}=(h_2, \ldots, h_M)$, with $h_i = |i \rangle \langle i |$, and 
\begin{equation}
    \rho_{\widehat{\vect{\phi}}} = e^{\ii \vect{h}\cdot \widehat{\vect{\phi}}} \margvsn e^{-\ii \vect{h}\cdot \widehat{\vect{\phi}}}.
\end{equation}
From the definition of $\Lambda$, \eqref{deflambda}, and the fact that   $\xi_k = n_k'$ at $\vect{\Phi}_{\alpha,\alpha}$, the quadratic expansion at that point  can be seen to involve only  the values  of the first derivatives of $\xi_k$   as well as the second derivatives with respect to  $\vect{\eta}$ of the function
\begin{equation}
    \mathcal{Z} := \sum_{k=1}^{M} \xi_k = \mathrm{tr} \left(  e^{{\ii}\vect{h}\cdot \vect{\eta}}  \margvs    \right),
\end{equation}
which can be seen to be $\widehat{\vect{\phi}}$-independent. Three quantities will be particularly important for our calculation. The first is the derivative tensor of  $\xi_k$ with respect to $\widehat{\vect{\phi}}$, which at the diagonal point $\vect{\Phi}_{\alpha,\alpha}$ coincides with the Jacobian matrix of the classical map \eqref{eq:map} at $\vect{\phi}_\alpha$:
\begin{equation}
 \left. \frac{\partial \xi_k}{\partial \widehat{\varphi}_l} \right|_{\alpha,\alpha}     =      \ii \langle k|U  \big[\, h_l\, \, ,\, \rho_{\widehat{\vect{\phi}}_\alpha}\, \big]  U^{\dagger}| k \rangle = \left. \frac{\partial \widetilde{\nu}_{\vect{\phi},k}}{\partial \phi_l}\right|_{\vect{\phi}_\alpha} .
 \end{equation}
 The other two quantities of interest are the first and  second derivative tensors of $\mathcal{Z}$ with respect to $\vect{\eta}$, which are 
 \begin{equation}
 \begin{aligned}
 \left. \frac{\partial {\mathcal Z}}{\partial \eta_l} \right|_{\alpha,\alpha}     =   \ii \langle l | \rho |l \rangle  & =    i \overline{n}_l , \\
 \left. \frac{\partial^2 {\mathcal Z}}{\partial \eta_l\partial \eta_m} \right|_{\alpha,\alpha}     =   -\langle l | \rho |l \rangle \delta_{lm} & =    - \overline{n}_l \delta_{lm}. 
 \end{aligned}
\end{equation}
In terms of the new variables and using the above results, the quadratic term in expansion \eqref{quadlambda2} reads
\begin{equation}
    \delta\vect{\Phi}^T  H_{\alpha, \alpha} \delta\vect{\Phi}  =  \sum_{i=2}^{M} \overline{n}_i \delta\eta_i^2 + \sum_{k=1}^{M}\frac{1}{\overline{n}_k'}\left( \delta \widehat{\vect{\phi}} \cdot \left. \frac{\partial \widetilde{\nu}_{\vect{\phi},k}'}{\partial \widehat{\vect{\phi}}}\right|_{\alpha} + \delta \vect{\eta} \cdot \left.\frac{\partial\xi_k}{\partial \vect{\eta}}\right|_{\alpha} \right)^2. 
\end{equation}
Next, using $\xi_1 = \mathcal{Z}- \sum_{k=2}^M \xi_k$, we rewrite the $k=1$ term of the second sum in terms of quantities involving $\xi_k$ for  $2 \leq k \leq M$:
\begin{equation}
\left.\frac{\partial \widetilde{\nu}_1}{\partial \widehat{\varphi}_l}\right|_{\alpha} =  - \sum_{k=2}^M \left.\frac{\partial \widetilde{\nu}_{\vect{\phi},k}}{\partial \widehat{\varphi}_l}\right|_{\alpha}, \qquad  \left.\frac{\partial \xi_1}{\partial \eta_l}\right|_{\alpha} =  \ii n_l - \sum_{k=2}^M \left.\frac{\partial \xi_k}{\partial \eta_il}\right|_{\alpha},
\end{equation}
and define the variables
\begin{equation}
    \zeta_k = \delta \widehat{\vect{\phi}}\cdot \left. \frac{\partial \widetilde{\nu}_{\vect{\phi},k}'}{\partial \widehat{\vect{\phi}}}\right|_{\widetilde{\vect{\phi}}_\alpha} + \delta \vect{\eta} \cdot \left.\frac{\partial\xi_k}{\partial \vect{\eta}}\right|_{\vect{\phi}_\alpha} +  \ii\,  \overline{n}_k' (\delta \vect{\eta}\cdot \ov{n})  
\end{equation}
for $k=2,\ldots,M$. Then the quadratic term can be written in terms of $M-1$-dimensional vectors and square matrices as
\begin{equation}
 \delta\vect{\Phi}^T  H_{\alpha, \alpha} \delta\vect{\Phi}  = \delta \vect{\eta}^T \vect{A}_{\ov{n}} \delta \vect{\eta} + {\vect{\zeta}}^T \vect{A}^{-1}_{\ov{n}'}{\vect{\zeta}},
 \end{equation}
where,
for any normalized distribution  $ \ov{n}$, we define $\vect{A}_{\ov{n}}$ and its inverse as the   $(M-1)\times(M-1)$ matrices  with elements
\begin{equation}
[\vect{A}_{\ov{n}}]_{i,j} = \overline{n}_i \delta_{i j} - \overline{n}_i \overline{n}_j,  \qquad [\vect{A}^{-1}_{\ov{n}}]_{ij}= \frac{\delta_{i j}}{n_i} + \frac{1}{n_1}, 
\end{equation}
with the understanding that  indices run from  $ 2$ to $M$. Implementing the change of variable    $\vect{\varphi} \rightarrow {\vect{\zeta}}$ and performing the resulting independent Gaussian integrals in $\widetilde{\vect{z}}$ and $\vect{\eta}$ yields
\begin{align}
     \int d \mu(\vect{\Phi}) e^{- \frac{1}{2}\delta\vect{\Phi}^T  H_{\alpha \alpha} \delta\vect{\Phi} } & = 
     \left| \det \frac{\partial \widetilde{\vect{\nu}}_{\vect{\phi}} }{\partial \vect{\phi} }\right|^{-1}_{\scriptscriptstyle \vect{\phi}_\alpha}\! \int\! \frac{d^{\scriptscriptstyle M\!-\!1}\!{\vect{\zeta}} }{(2 \pi)^{M-1}}\,  e^{- \frac{1}{2}\delta{\vect{\zeta}}^T  \vect{A}^{-1}_{\ov{n}'} \delta{\vect{\zeta}} } \int \frac{\!d^{\scriptscriptstyle M\!-\!1}\!\vect{\eta}}{(2 \pi)^{M-1}} e^{- \frac{1}{2}\delta\vect{\eta}^T  \vect{A}_{\ov{n}} \delta\vect{\eta} } \\
   & =  \frac{1}{(2 \pi)^{M-1} }\left| \det \frac{\partial \widetilde{\vect{\nu}}_{\vect{\phi}}}{\partial \vect{\phi}}\right|^{-1}_{\vect{\phi}_\alpha}  \sqrt{\frac{\det \vect{A}_{\ov{n}'}}{\det \vect{A}_{\ov{n}}}}.
\end{align}
Finally, factoring $\vect{A}_{\ov{n}}$ as
\begin{equation}
    \vect{A}_{\ov{n}} = \left[\sum_{i=2}^M \sqrt{ \overline{n}_i} | i \rangle \langle i | \right]  \left[\sum_{i,j=2}^M \left( \delta_{ij}-\sqrt{ \overline{n}_i\overline{n}_j}\right)|i\rangle \langle j|\right] \left[\sum_{i=2}^M \sqrt{ \overline{n}_i} | i \rangle \langle i | \right],
\end{equation}
it is straightforward to see that 
\begin{equation}
   \det \vect{A}_{\ov{n}} = \prod_{i=1}^{M} \overline{n}_i. 
\end{equation}
Comparing with \eqref{gaussintapp}, we obtain
\begin{equation}
    \det{H}_{\alpha, \alpha} = \left(\prod_{i=1}^{M}\frac{\overline{n}_i}{\overline{n}_i'}\right) \left|\det\frac{\partial{\widetilde{\vect{\nu}}_{\vect{\phi}}}}{\partial{\vect{\phi}}}\right|_{\vect{\phi}_\alpha}^2 .
\end{equation}

\section{}
\label{WKBappendix}

In this appendix we derive the semiclassical formula \eqref{eq:wkb} and the phase stated given in Eq.~\eqref{eq:action}. For the initial and final virtual states  $\ket{\widetilde{\vect{\nu}}_{\vect{\phi}},\widetilde{\vect{\phi}}_{\vect{\phi}}}$ and $ \ket{\vect{\nu},\vect{\phi}}$ let us define the complex amplitudes
\begin{equation}
z_i(\vect{\phi})
=\sqrt{\overline n_i}\,e^{\ii\phi_i}, \qquad \widetilde{z}_k(\vect{\phi}) = \sqrt{\widetilde{\nu}_{\vect{\phi},k}}\,
e^{\ii\widetilde{\phi}_k(\vect{\phi})},
\end{equation} so that 
$ 
\widetilde{z}_k(\vect{\phi})
=\sum_iU_{ki}z_i(\vect{\phi})$.
Then, writing $\sigma(\vect{\phi})$ as
\begin{equation}
\sigma(\vect{\phi})
=
-\ov{n}\cdot\vect{\phi}
-\ii\sum_k\overline n'_k
\ln\!\left(
\frac{\widetilde{z}_k(\vect{\phi})}{\sqrt{\overline n'_k}}
\right),
\label{eq:sigma}
\end{equation}
it follows that
\begin{equation}
\frac{\partial\sigma}{\partial\phi_i}
=
-\overline n_i+
z_i\sum_k\overline n'_k\frac{U_{ki}}{\widetilde{z}_k},
\label{eq:first-derivative}
\end{equation}
where we omit the dependence on $\vect{\phi}$ for simplicity. 
At a preimage $\vect{\phi}_\alpha$, $\overline n'_k=|\widetilde{z}_k|^2$, and hence $\overline n'_k/\widetilde{z}_k=\widetilde{z}_k^*$. Since $\widetilde{\vect{z}}=U\vect{z}$ and $U$ is unitary, $\sum_k\widetilde{z}_k^*U_{ki}=z_i^*$. Therefore, the derivative in
Eq.~\eqref{eq:first-derivative} is zero for every $\vect{\phi}_\alpha\in\vect{\phi}(\ov{n}')$.

Next, we define
\begin{equation}
g(\vect{\phi}):=-\ii\sigma(\vect{\phi}),
\qquad[G_\alpha]_{ij}:=\left.\frac{\partial^2g}
{\partial\phi_i\partial\phi_j}\right|_{\vect{\phi}_\alpha},
\qquad i,j=2,\ldots,M.
\end{equation}
From Eq.~\eqref{eq:sigma}
\begin{equation}
[G_\alpha]_{ij}=-\left.\sum_{\ell=1}^{M}\overline n'_\ell
\frac{\partial^2}{\partial\phi_i\partial\phi_j}
\ln \widetilde{z}_\ell\right|_{\vect{\phi}_\alpha}.
\label{G-second}
\end{equation}
To rewrite \eqref{G-second}, first note the identity
\begin{equation}
\sum_{\ell=1}^{M}
\widetilde{\nu}_{\vect{\phi},\ell}
\frac{\partial}{\partial\phi_j}
\ln \widetilde{z}_\ell=\sum_{\ell=1}^{M}
\widetilde{z}_\ell^*\frac{\partial \widetilde{z}_\ell}{\partial\phi_j}
=\vect{z}^{\dagger}
\frac{\partial\vect{z}}{\partial\phi_j}=\ii \overline n_j.
\end{equation}
Differentiating this identity with respect to $\phi_i$ gives
\begin{equation}
0=\sum_{\ell=1}^{M}
\frac{\partial \widetilde{\nu}_{\vect{\phi},\ell}}
{\partial \phi_i}\frac{\partial}{\partial \phi_j}
\ln \widetilde{z}_\ell+\sum_{\ell=1}^{M}
\widetilde{\nu}_{\vect{\phi},\ell}
\frac{\partial^2}{\partial \phi_i \partial \phi_j}\ln \widetilde{z}_\ell.
\end{equation}
At the stationary point, $\widetilde{\nu}_{\vect{\phi}_\alpha,\ell}=\overline n'_\ell$, so comparison with \eqref{G-second} yields
\begin{equation}
[G_\alpha]_{ij}
=\left.\sum_{\ell=1}^{M}
\frac{\partial
\widetilde{\nu}_{\vect{\phi},\ell}}
{\partial\phi_i}
\frac{\partial}{\partial\phi_j}
\ln \widetilde{z}_\ell\right|_{\vect{\phi}_\alpha}.
\end{equation}
Finally, because $\sum_{\ell}\widetilde{\nu}_{\vect{\phi},\ell}=1$,
we have $\sum_{\ell}\partial_{\phi_i}
\widetilde{\nu}_{\vect{\phi},\ell}=0$. Subtracting
$\partial_{\phi_j}\ln \widetilde{z}_1$ inside the sum gives

\begin{equation}
[G_\alpha]_{ij}
=
\left.
\sum_{k=2}^{M}
\frac{\partial\widetilde{\nu}_{\vect{\phi},k}}
{\partial\phi_i}
\frac{\partial}{\partial\phi_j}
\ln\!\left(\frac{\widetilde{z}_k}{\widetilde{z}_1}\right)
\right|_{\vect{\phi}_\alpha}.
\end{equation}
Therefore $G_\alpha=J_\alpha^TC_\alpha$, where $J_\alpha$ is the Jacobian of the classical map,
\begin{equation}
[J_\alpha]_{ki}
=
\left.
\frac{\partial\widetilde{\nu}_{\vect{\phi},k}}
{\partial\phi_i}
\right|_{\vect{\phi}_\alpha},
\qquad
[C_\alpha]_{kj}
=
\ii\,z_j\left(
\frac{U_{kj}}{\widetilde{z}_k}-\frac{U_{1j}}{\widetilde{z}_1}
\right)_{\vect{\phi}_\alpha},
\end{equation}
for $i,j,k=2,\ldots,M$.
To evaluate $\det C_\alpha$, define
$B_{kj}=U_{kj}-(\widetilde{z}_k/\widetilde{z}_1)U_{1j}$. Factoring $\ii z_j$ out of column $j$ and $\widetilde{z}_k^{-1}$ out of row $k$, for $j,k=2,\ldots,M$, gives
\begin{equation}
\det C_\alpha
=
\ii^{M-1}\frac{\prod_{j=2}^{M}z_j}{\prod_{k=2}^{M}\widetilde{z}_k}\det B.
\end{equation}
If the first column of $U$ is replaced by $\widetilde{\vect{z}}=U\vect{z}$, the determinant of the resulting matrix is both $z_1\det U$ and $\widetilde{z}_1\det B$, by the Schur-complement formula. Hence $\det B=(z_1/\widetilde{z}_1)\det U$, and
\begin{equation}
\det(G_\alpha)
=
\ii^{M-1}\det J_\alpha
\frac{\prod_i z_i(\vect{\phi}_\alpha)}
{\prod_k\widetilde{z}_k(\vect{\phi}_\alpha)}
\det U.
\end{equation}
At the saddle, $z_i(\vect{\phi}_\alpha)=\sqrt{\overline n_i}e^{\ii\phi_i}$ and
$\widetilde{z}_k(\vect{\phi}_\alpha)=\sqrt{\overline n'_k}e^{\ii\widetilde{\phi}_k}$. Consequently
\begin{equation}
|\det(G_\alpha)|
=
|\det J_\alpha|
\sqrt{\frac{\prod_i\overline n_i}
{\prod_k\overline n'_k}},
\end{equation}
and
\begin{align}
\delta_\alpha
:={}&-\frac{1}{2}\arg(\det G_\alpha)
\\={}&-\frac{(M-1)\pi}{4}
-\frac{1}{2}\arg(\det U) -\frac{1}{2}\arg(\det J_\alpha)+\frac{1}{2}
\left(\sum_{k=1}^{M}\widetilde{\phi}_{k,\alpha}
-\sum_{i=1}^{M}\phi_{i,\alpha}
\right).
\end{align}
Here, $\operatorname{arg}$ denotes the principal argument in
$(-\pi,\pi]$. The same branch is used for both determinants.
While $\operatorname{arg}(\det U)$ is generally arbitrary, the fact that $\det J_\alpha$ is real and nonzero implies
\begin{equation}
\arg(\det J_\alpha)
=\begin{cases}
0, & \det J_\alpha>0,\\
\pi, & \det J_\alpha<0,
\end{cases}
\end{equation}
so that
\begin{equation}
-\frac{1}{2}\arg(\det J_\alpha)
=\left[
\operatorname{sgn}(\det J_\alpha)-1
\right]\frac{\pi}{4}.
\end{equation}

\section*{References}
\bibliographystyle{iopart-num}
\bibliography{references_clean_etal}
\end{document}